\documentclass[onecolumn,authoryear]{els-mrw} 

\usepackage{amsmath,amssymb,amsfonts,amsthm,makeidx,graphicx}
\usepackage[varg]{txfonts}
\usepackage{helvet, multirow}

\renewcommand{\vec}[1]{\boldsymbol{#1}}

\newcommand*\jpcenterstack[2]{%
  \ensuremath{%
    \mathrel{%
      \mathchoice{%
        \vcenter{%
          \offinterlineskip
          \halign{\hfil$\displaystyle##$\hfil\cr#1\cr#2\cr}%
        }%
      }
      {%
        \vcenter{%
          \offinterlineskip
          \halign{\hfil$\textstyle##$\hfil\cr#1\cr#2\cr}%
        }%
      }
      {%
        \vcenter{%
          \offinterlineskip
          \halign{\hfil$\scriptstyle##$\hfil\cr#1\cr#2\cr}%
        }%
      }
      {%
        \vcenter{%
          \offinterlineskip
          \halign{\hfil$\scriptscriptstyle##$\hfil\cr#1\cr#2\cr}%
        }%
      }%
    }%
  }%
}
\newcommand*\la{\jpcenterstack{<}{\sim}}
\newcommand*\ga{\jpcenterstack{>}{\sim}}

\newcommand{\msun}{\ensuremath{M_{\odot}}}
\newcommand{\lsun}{\ensuremath{L_{\odot}}}
\newcommand{\rsun}{\ensuremath{R_{\odot}}}

\newcommand{\Teff}{\ensuremath{T_{\rm eff}}}

\newcommand{\mdot}{\ensuremath{\dot{M}}}

\newcommand{\msunyr}{\ensuremath{M_{\odot} {\rm yr}^{-1}}}

\newcommand{\beq}{\begin{equation}}
\newcommand{\eeq}{\end{equation}}
\newcommand{\beqa}{\begin{eqnarray}}
\newcommand{\eeqa}{\end{eqnarray}}

\newcommand{\nbeq}{\begin{equation*}}
\newcommand{\neeq}{\end{equation*}}

\newcommand{\kms}{\ensuremath{{\rm km}\,{\rm s}^{-1}}}

\newcommand{\dd}{{\rm d}}

\newcommand{\HI}  {H\,{\sc i}}

\newcommand{\HeI} {He\,{\sc i}}
\newcommand{\HeII}{He\,{\sc ii}}

\newcommand{\NIV}{N\,{\sc iv}}
\newcommand{\NV}{N\,{\sc v}}

\newcommand{\CaI}{Ca\,{\sc i}}
\newcommand{\CaII}{Ca\,{\sc ii}}

\newcommand{\FeI}{Fe\,{\sc i}}

\newcommand{\FeXVII}{Fe\,{\sc xvii}}

\newcommand{\Hd} {H$_{\rm \delta}$}
\newcommand{\Hg} {H$_{\rm \gamma}$}

\newcommand{\Ha} {H$_{\rm \alpha}$}

\newcommand{\Rstar}{\ensuremath{R_{\ast}}}

\newcommand{\Mstar}{\ensuremath{M_{\ast}}}

\newcommand{\logg}{\ensuremath{\log g}}

\newcommand{\vth}{\ensuremath{v_{\rm th}}}

\newcommand{\grad}{\ensuremath{g_{\rm rad}}}

\def\aj{AJ}%
\def\araa{ARA\&A}%
\def\apj{ApJ}%
\def\apjl{ApJ}%
\def\apjs{ApJS}%
\def\aap{A\&A}%
\def\aapr{A\&A~Rev.}%
\def\mnras{MNRAS}%
\def\pra{Phys.~Rev.~A}%
\def\solphys{Sol.~Phys.}%
\def\na{New Astronomy}%
\def\memsai{Mem.~Soc.~Astron.~Italiana}%
\begin{document}

\chapter{Stellar Atmospheres}\label{chap1}

\author[1]{Joachim Puls}
\author[2,3]{Artemio Herrero}
\author[3,2]{Carlos Allende Prieto}

\address[1]{\orgname{LMU Munich}, \orgdiv{University Observatory},
\orgaddress{Scheinerstr. 1, D-81679 M\"unchen, Germany}}
\address[2]{\orgname{Universidad de La Laguna}, \orgdiv{Departamento
de Astrof\'isica}, \orgaddress{Avda. Astr. Francisco S\'anchez, 
E-38206 La Laguna, Spain}}
\address[3]{\orgname{Instituto de Astrof\'isica de Canarias},
\orgaddress{C/ Via L\'actea s/n, E-38206 La Laguna, Spain}}

\articletag{Chapter Article tagline: update of previous edition,, reprint..}

\maketitle

\begin{glossary}[Glossary]

\term{corona} outermost, extended region of a (low-mass) star such as
the Sun, with very high temperatures (millions of Kelvin) and low densities.

\term{chromosphere} transition zone between photosphere and corona.

\term{photosphere} outer region of a star where most of the optical
light (and from other wavelengths) originates from. Geometrically thin
compared to the stellar radius, with temperatures on the order of the
effective one.

\term{quantitative spectroscopy} derivation of stellar atmospheric
parameters, incl. chemical surface abundances, by means of comparing
observed and synthetic spectra.

\term{specific intensity} basic quantity in radiative transfer,
proportional to transported energy, direction-dependent.

\term{spectral lines} narrow absorption or emission
features in a spectrum, resulting from electronic transitions between
two bound levels.

\term{stellar atmosphere} outer region of a star where all emitted 
light is produced. Comprises photosphere, and, if present,
chromosphere, corona, and wind.

\term{stellar wind} mass outflow from a star.

\end{glossary}
\begin{glossary}[Nomenclature]
\begin{tabular}{@{}lp{34pc}@{}}
CMF & comoving frame\\
LTE & local thermodynamic equilibrium\\
%MLT & mixing length theory\\
NLTE & non-LTE, or kinetic equilibrium\\
RE & radiative equilibrium\\
RT & radiative transfer\\
RMHD & radiation magneto-hydrodynamics\\
SED & spectral energy distribution\\
SN, SNe &supernova(e)\\
TE & thermodynamic equilibrium\\
UV & ultraviolet\\ 
\end{tabular}
\end{glossary}

\begin{abstract}
[Stars play a decisive role in our Universe, from its beginning
throughout its complete evolution. For a thorough understanding of
their properties, evolution, and physics of their outer envelopes,
stellar spectra need to be analyzed by comparison with numerical
models of their atmospheres. We discuss the foundations of how to
calculate such models (in particular, density and temperature
stratification, affected by convective energy transport in low-mass
stars), which requires a parallel treatment of hydrodynamics,
thermodynamics and radiative transfer. We stress the impact of
emissivities, opacities, and particularly their ratio (source
function), and summarize how these quantities are calculated, either
adopting or relaxing the assumption of LTE (local thermodynamic
equilibrium). Subsequently, we discuss the influence and physics of
stellar winds (and their various driving mechanisms as a function of
stellar type), rotation, magnetic fields, inhomogeneities, and
multiplicity. Finally, we outline the basics of quantitative
spectroscopy, namely how to analyze observed spectra in practice.] 
\end{abstract}

\paragraph{Key Points}
\begin{itemize}
\item The theory of stellar atmospheres provides us with physical 
models for the outermost stellar envelopes, and quantifies, in
dependence of atmospheric parameters, the run of, e.g., 
density, velocity, temperature, radiation field,
ionization/excitation of atoms, ions, molecules.
\item Such models base on a radiation (magneto-)hydrodynamic
and thermodynamic description, either simplified (one-dimensional,
static, steady-state), or, if required, accounting for convection, 
outflows, rotation, magnetic fields, and inhomogeneities (partly
multi-D).
\item Atmospheric models are utilized to synthesize the emergent SEDs,
that will be compared with observations to infer the stellar (and, if
present, wind) parameters, as well as the chemical surface composition.
\item The ionization/excition balance of the atmospheric atoms, ions, and
molecules is calculated either under the assumption of LTE, or, if the
impact of the radiation field dominates over collisional processes,
accounting for deviations from LTE (= non-LTE).
\item Stellar outflows (= winds) modify the outermost atmospheres and the
emergent SEDs. The majority of stellar winds are accelerated either by
gas pressure forces (when a hot corona is present), or by radiation
pressure forces (for luminous and/or hot stars).
\end{itemize}

\section{Introduction}\label{intro} 
Most of the information we collect from objects and regions beyond the
Solar System is carried by photons, and most of these photons are
generated in stellar interiors and transported throughout the stars.
In their journey to the outer stellar layers (atmospheres), they
change their properties and finally escape, after a last interaction
with the atmospheric atoms and molecules. Leaving the star behind,
photons travel to us, sometimes suffering additional interactions in
the nearly void interstellar medium, until they reach our instruments.

This information has to be interpreted with models describing the
physical processes that photons have been subject to. With this
formulation, we can recover the physical conditions that generated the
observed photons, and gather required knowledge about the conditions
inside stellar envelopes (temperatures, densities, pressures, chemical
composition, etc.) and also about the properties of stellar interiors.

Every photon counts for the analysis, but it is their spectral energy
distribution (SED) that contains the record of the physical processes
involved. Thus, the better this distribution is known, the better we
can describe those physical processes. A condensed description of this
distribution (usually called {\it photometric} information) provides
the basic information about the star, but describing its detailed
distribution (the {\it spectroscopic} information) requires an
accurate formulation of the physical laws governing stellar
atmospheres. The process of extracting the physical quantities and
magnitudes present in the system by analyzing the observed detailed
distribution of photons is what we call {\it quantitative
spectroscopy}.

The theories of radiative transfer and stellar structure are
continously tested and challenged through quantitative spectral
analyses of stellar atmospheres, both in individual systems and in
increasingly larger stellar samples. Joint studies of stars in the
same cluster or association, or of similar nature (like Blue or Red
Supergiants, high-mass X-ray binaries or extremely metal poor stars,
to mention only few) test and extend our knowledge of the theory of
stellar formation, structure and evolution, and inform us about the
fate of the different stars and stellar systems. This way they provide
us with quantitative information about the mechanical and radiative
energy injected into their environments and host galaxies, including
ionizing fluxes and newly synthesized elements. Combining this
information with observations of distant galaxies and very old stars,
we can build a scenario of cosmic evolution, also via exploiting the
information contained in the observation of extreme events, like
supernovae (SNe), gamma-ray bursts and black hole mergers.

All this knowledge relies on the accuracy of analyses of the light
emitted by stellar atmospheres: hot plasmas with strong physical
gradients kept bound by the action of gravity, but subject to
rotation, magnetic fields, radiation pressure, and
(radiation-magneto)hydrodynamic instabilities giving rise to spatial
inhomogeneities. Within this framework, we have to describe the
interaction of photons and matter at atomic and molecular level.

This is the subject of the theory of stellar atmospheres, and has to
invoke a variety of building blocks from our physical theories. Many
authors have contributed to its development and current status (see
the introductory chapters in classical (text)books{\footnote{for the
books, we cite the last editions, and we note that \cite{Payne1925} is
a PhD thesis}} like \cite{Payne1925}, \cite{Unsold1955},
\cite{Aller1963}, \cite{Mihalas1978}, \cite{Gray2021}, and
\citet[H\&M]{Hubeny2015}, rendering the Theory of Stellar Atmospheres
a corner stone for our knowledge of the Universe. A brief exposure of
the foundations of this theory is the purpose of the present chapter.

\section{Basic Considerations}\label{basics} 
In this section, we will introduce the major assumptions, definitions,
and techniques required to model stellar atmospheres. Since these will
strongly depend on the considered stellar type (cool vs. hot, compact
vs. extended) and the absence or presence of a mass outflow (wind), a
variety of concepts needs to be discussed. A particular difficulty
arises since in almost all cases the coupling of matter with the
radiation field must be accounted for, which requires accounting for
radiative transfer (RT), and increases significantly the computational
effort.

\subsection{Hydrostatic and Hydrodynamic Structure}
\label{hydro}
We begin with a description of the density structure. Under typical
conditions (except for coronae), atmospheric plasmas behave as ideal
gases, and (kinematic) viscosity terms are small (e.g.,
\citealt{Cowley1990}) and mostly negligible. In this case, the
hydrodynamic equations of continuity (i) and momentum (ii, ``Euler
equation'') read (with boldface indicating vectors)
\beq
\label{eq_hydro}
{\rm (i)}\quad\frac{\partial \rho}{\partial t}+\vec{\nabla} \cdot (\rho \vec{v})=0,
\qquad
{\rm (ii)}\quad\frac{\partial \rho \vec{v}}{\partial t}+\vec{\nabla} 
\cdot (\rho \vec{v} \otimes \vec{v})= -\vec{\nabla} p+\rho
\vec{g^{\rm ext}},
\eeq
with density $\rho$, velocity $\vec{v}$, gas pressure $p$ and external
accelerations $\vec{g^{\rm ext}}$, and $\otimes$ the dyadic (or outer)
product. In many cases, atmospheres are
adopted as stationary (steady-state, time-independent), and a 1-D
approximation is applied (but see below), either with plane-parallel
symmetry and height coordinate $z$ (if the extent of the atmosphere is
small compared to the stellar radius, \Rstar), or with spherical
symmetry and radial coordinate $r$ (if the extension is large, often
because of the presence of a wind or very low gravity). Furthermore,
if velocity fields can be neglected, the Euler equation collapses to
the equation of \textit{hydrostatic equilibrium}, $\dd p/\dd z = \rho
g^{\rm ext}$, stating that the pressure force needs to be balanced by
external forces. Under simplifying assumptions (neglect of radiative
acceleration, $T(r) \approx T_{\rm phot}$ with $T_{\rm phot}$ a
representative value), this equation can be solved
analytically\footnote{for an exact solution, numerical methods need to
be invoked} by means of the ideal gas equation of state, $P = N k_{\rm
B} T$ with particle density $N$, Boltzmann constant $k_{\rm B}$ and
temperature $T$. The solution is the well-known \textit{barometric
formula}, 
\beq 
\label{eq_scaleheight}
P(z)=P(z_0) \exp\,\Bigl(-\frac{(z-z_0)}{H}\Bigr) \quad \text{or} \quad
\rho(z)=\rho_(z_0) \exp\,\Bigl(-\frac{(z-z_0)}{H}\Bigr) \qquad
\text{with} \qquad
H=\frac{k_{\rm B} T_{\rm phot}}{\mu m_{\rm H} g_\ast}=
2\frac{v_{\rm s}^2}{v_{\rm esc}^2} \Rstar, \qquad
g_\ast = \frac{G \Mstar}{\Rstar^2}
\eeq
where $H$ is the (photospheric) \textit{scale height}, $\mu$ the mean
molecular weight in units of hydrogen atomic mass $m_{\rm H}$, $v_{\rm
s} = \sqrt{k_{\rm B} T/(\mu m_{\rm H})}$ the isothermal speed of
sound, $v_{\rm esc} = \sqrt{2 g_\ast \Rstar}$ the photospheric escape
velocity, and $g_\ast$ the photospheric gravitational acceleration,
with gravitational constant $G$. For the sun, the scale height is
roughly 150 km, for a typical O-dwarf with a surface temperature of
35,000 K it is 4500 km, while for a Red Supergiant it easily exceeds
10$^6$ km. 

If we now include the presence of an outwards directed velocity field
($v(r) > 0$), and still require stationarity, a spherically symmetric 
1-D description predicts the presence of a mass-flux, a
\textit{stellar wind}\footnote{for the case of spherical, so-called
Bondi-accretion with $v(r) < 0$ we refer to the literature, e.g.,
\citealt{Bondi1952}},
\beq
\label{eq_motion}
{\rm (i)}\quad r^2 \rho v(r) = {\rm const} = \frac{\mdot}{4\pi}, 
\qquad
{\rm (ii)}\quad \rho v(r) \frac{\dd v}{\dd r} = - \frac{\dd p}{\dd r} +
\rho g^{\rm ext}(r),
\eeq
with $\mdot$ the stellar mass-loss rate. The left-hand side (lhs) of
the \textit{equation of motion}, Eq. 3(ii), is the advection term,
resulting from inertia. A detailed comparison of the hydrostatic and
hydrodynamic description reveals that both formulations become similar
for low velocities $v(r) \ll v_{\rm s}$, i.e., that also in case of a
stellar wind the deep photosphere becomes quasi-hydrostatic, though
with a non-zero velocity controlled by Eq.~\ref{eq_motion}\,(i).

\begin{figure}[t]
\centering
\includegraphics[width=.6\textwidth, angle=180]{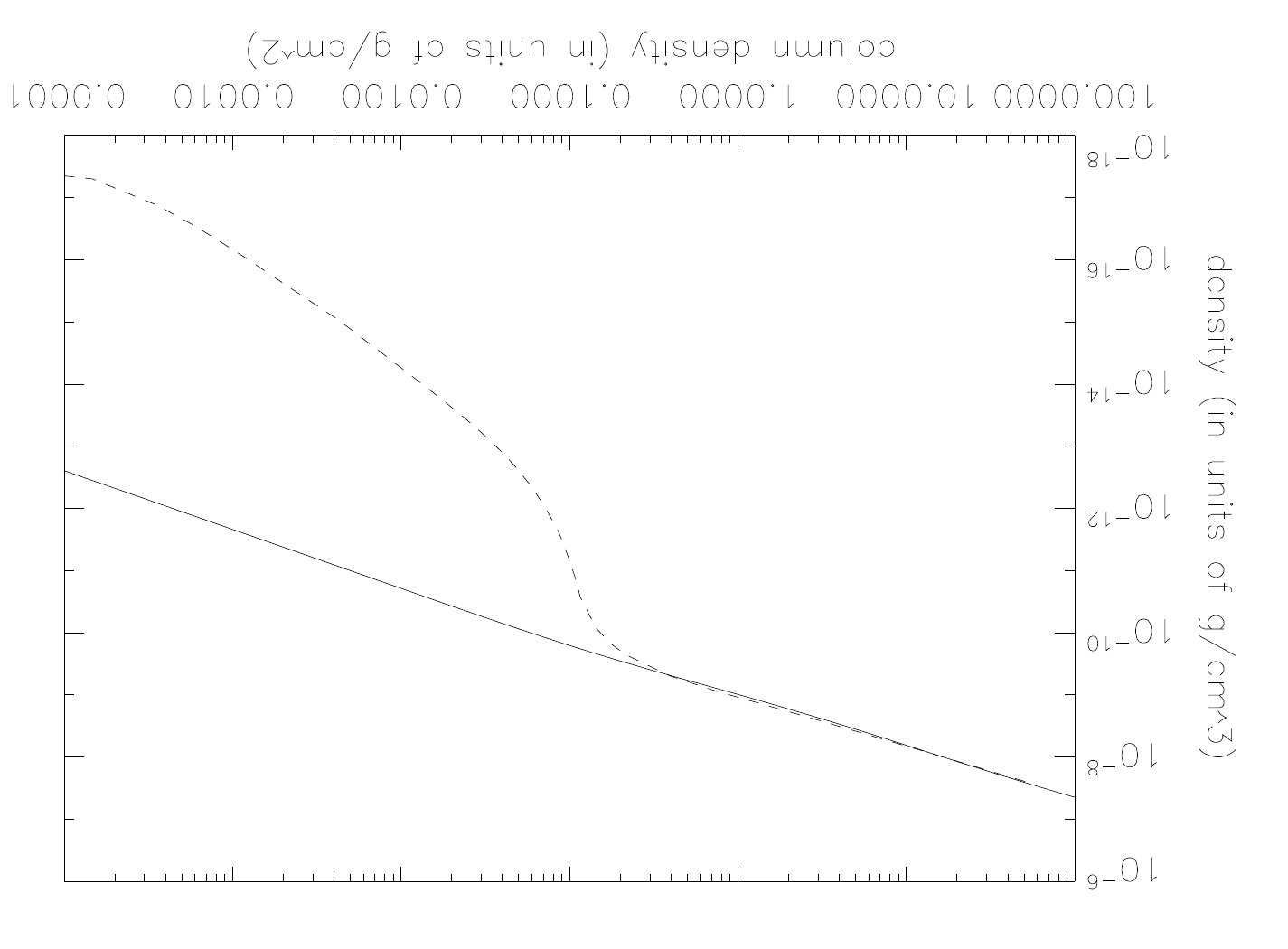}
\caption
{Density stratification as a function of column density ($\dd m = -
\rho \dd z$ or $\dd m = - \rho \dd r$), for an O-star with \Teff\ =
35,000~K and \logg\ = 4.0 (cgs-units). Solid: hydrostatic
stratification from TLUSTY (see Sect.~\ref{codes}), dashed:
hydrodynamic stratification with a wind of intermediate strength from
FASTWIND. When plotted as a function of $m$ (or optical depth $\tau$),
the outer densities (low column densities) for a wind-model are lower
than for a hydrostatic one. Since the cores of strong lines form
already at $m > 0.01$, in this case the wind would need to be
considered in an analysis (see Sect.~\ref{winds}).} 
\label{comprho}
\end{figure}

Stationary 1-D atmospheric models including both photosphere and wind
can be constructed in two ways. One possibility consists of 
integrating the equation of motion by accounting for all relevant
external accelerations\footnote{for solar-type winds, the additional
solution of the energy equation is required}, detailed in
Sect.~\ref{winds}. Approximate models (see Fig.~\ref{comprho}), on the
other hand, smoothly connect a (quasi-)hydrostatic subsonic solution
with a prescribed wind-velocity law and an input mass-loss rate, such
that the innermost velocities are calculated from $\mdot$ and
$\rho(r)$, whereas the wind densities result from $\mdot$ and $v(r)$.
%fig comprho

Because of increasing computational power, a variety of
multi-dimensional (multi-D) atmospheric models have been considered. 
For cooler type stars, this is mostly to investigate the effects of
convection and pulsations, and for hotter type stars to investigate
the role of inhomogeneities and rotation in their winds, while the
impact of magnetic fields has to be considered in both cases.

\subsection{Radiative Transfer (RT)}
\label{rt}
As already mentioned, various radiation field related quantities need
to be known to calculate model atmospheres and corresponding synthetic
spectra. The fundamental quantity is the \textit{specific intensity}
$I(\vec{r},\vec{n}, \nu, t)$ being the radiation energy with
frequencies within $(\nu, \nu + \dd \nu)$, which is transported
through a projected area element $\dd \sigma \cos \theta$ into
direction $\vec{n}$ (with $\theta$ the angle between the surface normal
and the propagation direction), per time interval $\dd t$ and solid angle
$\dd \Omega$, such that the transported energy is given by
\beq
\label{eq_Inu}
\dd E = I(\vec{r},\vec{n}, \nu, t) \cos \theta \dd \sigma \dd \nu
\dd t \dd \Omega.
\eeq
By definition, the specific intensity remains constant between an
emitting and receiving area (e.g., a telescope), as long as there is
no absorption and emission in between. Otherwise, and in the absence
of general relativistic effects\footnote{for the inclusion of GR
effects, see, e.g., \citet{Younsi2012}} (i.e., photons move on
straight lines at constant frequency), the equation of radiative
transport (RT) reads
\beq
\label{eq_RT}
\Bigl(\frac{1}{c}\frac{\partial}{\partial t}+\vec{n}\cdot\vec{\nabla}\Bigr)
I_\nu = \eta_\nu - \chi_\nu I_\nu,
\eeq
where the arguments ($\vec{r},{\vec n},t$) have been omitted.
$\eta_\nu$ is the macroscopic emissivity, and $\eta_\nu \dd s$ is the
specific intensity added by emission along a path-length $\dd s$ (with
$\vec{n}\cdot\vec{\nabla} = \dd/\dd s)$. $\chi_\nu$ is the macroscopic
opacity, such that $\chi_\nu I_\nu \dd s$ is the specific intensity
removed by absorption or scattering. Further details will be discussed
in Sect.~\ref{opa_eta}.

Because of the directional derivative and without any specific symmetry, 
two directional angles have to be accounted for. If non-Cartesian
coordinates were used (e.g., spherical ones), these angles change with
location, and the above equation becomes a partial differential
equation with six(!) independent variables, to be solved for each
frequency $\nu$. This immediately reveals that any
radiation-hydrodynamics problem suffers from a bottleneck established
by RT, and approximate methods are required for reasonable turnaround times.

After solving the RT equation for $I_\nu$ (in most cases,
numerically), one can integrate the specific intensity over $\dd
\Omega/(4 \pi)$, over $\vec{n}\, \dd \Omega$, and over $\vec{n} \vec{n}\,
\dd \Omega$ to obtain the mean intensity, $J_\nu $, the radiative flux,
$\vec{\mathcal{F}}_\nu$ and the radiation stress tensor,
$\vec{P}_\nu$, respectively, which are inevitable quantities when
considering, e.g., scattering and photoionization, 
radiative acceleration, and radiative pressure.

\begin{figure}[t]
\centering
\includegraphics[width=.6\textwidth]{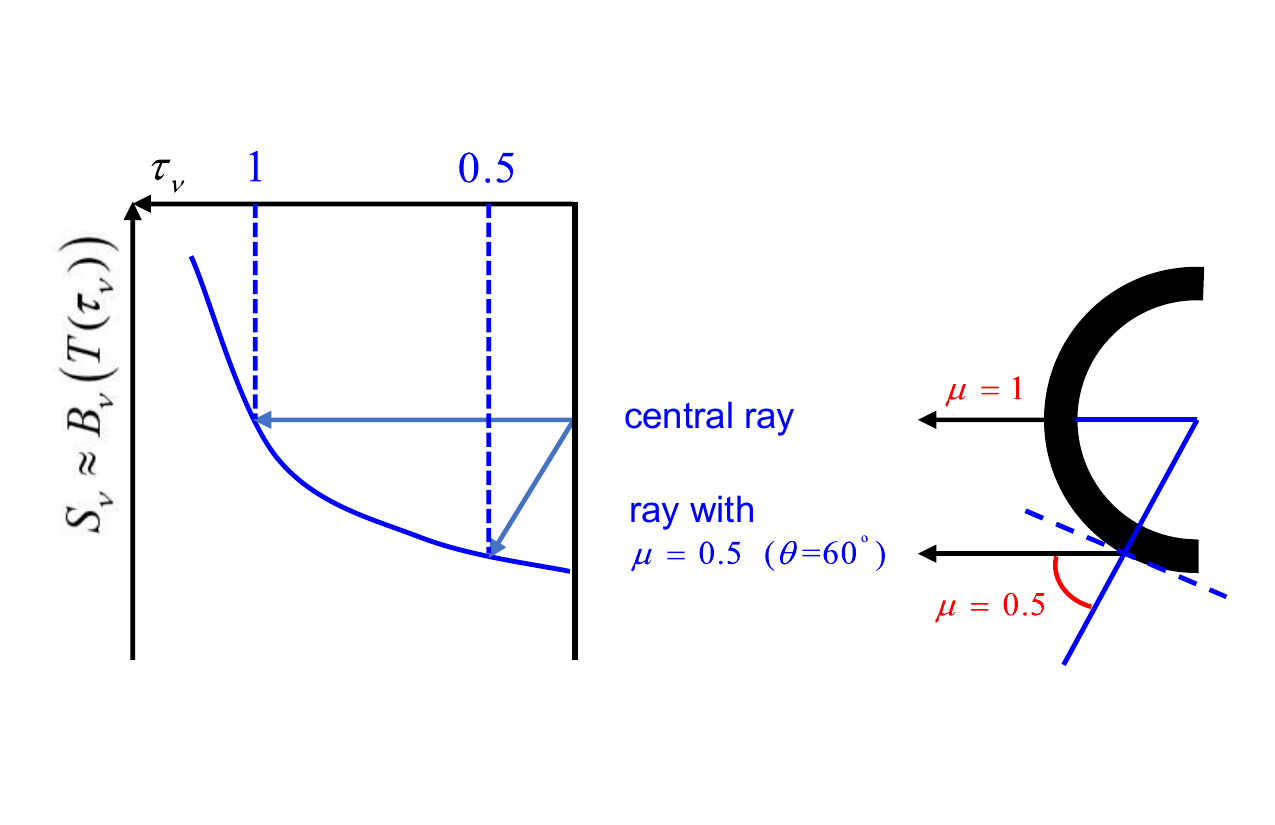}
\caption{Sketch of limb darkening. Because of an outwards decreasing 
source-function (see text), the emergent intensity decreases from the
center to the stellar limb (cf. Eq.~\ref{eq_EddBarb}).}
\label{limb_sketch}
\end{figure}

Assuming stationarity and either (i) plane-parallel or (ii) spherical
geometry, the number of independent variables becomes significantly
reduced, and the corresponding equations read
\beq
\label{eq_RT_1D}
{\rm (i)}\quad
\mu \frac{\dd}{\dd z} I_\nu(z,\mu) = \eta_\nu - \chi_\nu I_\nu,
\qquad
{\rm (ii)}\quad
\Bigl(\mu \frac{\partial}{\partial r} + \frac{1-\mu^2}{r}
\frac{\partial}{\partial \mu}\Bigr) 
I_\nu(r,\mu) = \eta_\nu - \chi_\nu I_\nu.
\eeq
where $\mu = \cos \theta$. A clever way to get rid of the
$\partial/\partial \mu$ derivative in Eq.~\ref{eq_RT_1D}\,(ii) is to
use a so-called \textit{p-z geometry} (\citealt{HR1971}, see also
\citealt{Crivellari2019}, Chapter 5), where the RT equation is solved
along impact parameters $p$ for ``height'' coordinates $z$, and the
directional derivative from Eq.~\ref{eq_RT} collapses to $\dd/\dd
z|_p$ along $p$ = const.  Often, the dimensionless \textit{optical
depth} (here for (i) Cartesian and (ii) spherical coordinates) 
\beq 
\label{eq_opticaldepth}
{\rm (i)}\quad \dd \tau_\nu = - \chi_\nu \dd z, 
\qquad \tau_\nu(z) = \int_{z}^{\infty} \chi_\nu(z) \dd z, 
\qquad 
{\rm (ii)}\quad \dd
\tau_\nu = - \chi_\nu \dd r, 
\qquad \tau_\nu(r) = \int_{r}^{\infty} \chi_\nu(r) \dd r, 
\eeq
is introduced, where $\tau_\nu(\infty) = 0$ at the outer boundary. In both cases, $\tau_\nu$
is better suited to act as a spatial ``coordinate'' than $z$ or $r$.
Dividing Eq.~\ref{eq_RT_1D}\,(i) by $(-\chi_\nu)$, we obtain (in
plane-parallel symmetry)
\beq
\label{eq_RT_pp}
\mu\frac{\dd I_\nu (z,\mu)}{\dd \tau_\nu} = I_\nu - S_\nu, 
\qquad S_\nu = \frac{\eta_\nu}{\chi_\nu},
\qquad S_\nu^{\rm TE} = B_\nu(T) = \frac{2 h \nu^3}{c^2}\frac{1}
{{\rm e}^{\frac{h\nu}{k_{\rm B} T}}-1},
\eeq
with \textit{source function} $S_\nu$. The \textit{Kirchhoff-Planck
law} states that in thermodynamic equilibrium (TE) the source function
is given by the Planck function ($h$ is the Planck constant), i.e.,
that under TE conditions the emissivity can be simply calculated from
the opacity and temperature. Eq.~\ref{eq_RT_pp}\,(left) can be easily
integrated, and for a ``semi-infinite'' atmosphere (with $\tau_\nu=
(0, \infty)$ at the surface and bottom of the atmosphere,
respectively, and $I_\nu(\tau_\nu) \exp(-\tau_\nu) \rightarrow 0$ for
$\tau_\nu \rightarrow \infty$), the specific intensity at the outer
boundary and $\mu > 0$ (the so-called \textit{emergent intensity})
results as 
\beq
\label{eq_solRT_pp}
I_\nu^{\rm em}(\mu) = I_\nu(\tau_\nu=0,\mu>0) = \int^\infty_{0} S_\nu(t)
{\rm e}^{-(t/\mu)} \frac{\dd t}{\mu}.
\eeq
This solution -- the emergent intensity is just the Laplace-transform
of the source function -- shows clearly the major role of the latter
in atmospheric calculations where the lower atmosphere is typically
located at large optically depths (high densities, $\tau_\nu \gg 1$). 
It is thus sufficient to know $S_\nu$ as function of optical depth to
solve for the emergent intensity. The importance of $S_\nu$ becomes
also evident when solving Eq.~\ref{eq_solRT_pp} by approximating the
source function to be linear in $\tau_\nu$, which results in the
so-called \textit{Eddington-Barbier relation}, 
\beq
\label{eq_EddBarb}
I_\nu^{\rm em}(\mu) \approx S_\nu(\tau_\nu = \mu).
\eeq
This relation states that, e.g., for vertical rays ($\mu = 1$) ``we
see'' just the source-function at $\tau_\nu = 1$. In particular, it
explains stellar limb-darkening of photospheres: for decreasing $\mu$
(center to limb), the specific intensity decreases, since the source
function (evaluated at $\tau_\nu=\mu$) usually decreases outwards with
height (e.g., in the simple case that $S_\nu \approx B_\nu(T)$ with
outwards decreasing $T$, see Fig.~\ref{limb_sketch}). As well, the
above relation explains the absorption seen in spectral lines, since
their opacity is larger than in the neighbouring continuum, i.e.,
$\tau_\nu = 1$ is reached further out in the photosphere, implying a
lower source-function and thus lower intensities in the lines.
%fig_limb

One major problem might appear when including velocity fields into the
RT equation, In principle, all opacities and emissivities need to be
evaluated at the comoving (= atomic) frame frequency (CMF), i.e.,
accounting for Doppler-shifts between the inertial, observer's
rest-frame and the comoving material, $\nu_{\rm CMF} \approx \nu_{\rm
obs}(1- \vec{n}\cdot \vec{v}/c)$ in the non-relativistic approximation
(note the projection!). For comparatively low velocities (e.g., winds
from Red Supergiants) this is usually not a problem. For superthermal
speeds ($v > \vth$, typical for winds from hot massive stars),
however, a very high radial and often also angular grid resolution
would be required,
%together with frequencies extending between
%$[v_{\rm max}, -v_{\rm max}]$ (in velocity space) centered around one
%specific line. 
to resolve the transfer in the most relevant line-cores, which are
Doppler-broadened with a width corresponding to the thermal velocity
(see Sect.~\ref{lines}). In effect, a solution in the observer's frame
becomes very time-consuming.  There are various possibilities to
circumvent this problem. Often, the RT equation is solved in the CMF,
which however is only advantegeous when the velocity field is
monotonic. For details regarding the spherical symmetric\footnote{for
a summary, see \citealt{Crivellari2019}, Chap. 5} and the general case
(multi-D, relativistic), we refer to H\&M and references therein.

An alternative approach is given by the so-called \textit{Sobolev
theory} \citep{Sobolev1960}, which aims at an analytical solution of
the line-transfer problem in rapidly expanding atmospheres. This
approximation makes advantage of the fact that because of the low
thermal speed compared to the bulk velocities, the geometrical extent
of a zone where stellar radiation can interact with a line-transition
from a moving ion (the so-called resonance zone) is quite narrow, of
order $L_{\rm Sob} (r)= v_{\rm th}/(\dd v/\dd r)$. In the original
approach, all quantities except for the velocity gradient are then
approximated as being constant within such a resonance zone. The
line-profile weighted and frequency integrated specific intensity,
$\bar I$, which is central to many applications (line source function
and radiative line acceleration) then becomes a purely local quantity,
and can be easily calculated. An excellent presentation of the
complete approach has been given by \citet{RybickiHummer1978}, and for
a summary of how to include additional effects we refer to
\citet[Chap. 5]{Crivellari2019}. 

\subsection{Energy Transport}
\label{energy}
In order to derive the atmospheric hydrodynamic structure, and the
opacities and emissivites controlling the radiation field
($\rightarrow$ radiatitive acceleration \grad), we need to know the
temperature stratification (under TE conditions, this would provide us
with the source function). The temperature stratification results from
the total (matter + radiation) energy equation, which in stationary
and static ($\vec{v}=0$) atmospheres (for time-dependent and/or
hydrodynamic conditions, see H\&M) 
collapses to 
\beq
\label{eq_energy_stat}
\vec{\nabla} \cdot [\mathcal{F} + \vec{F}^{\rm c} + \vec{F}^{\rm
conv} + {\ldots} ] = 0, \qquad \text{i.e., either } \sum F = \text{const
(plane-parallel sym.) or} \qquad
r^2 \sum F = \text{const (spherical sym.)}.
\eeq
$\vec{F}^{\rm c}$ and $\vec{F}^{\rm conv}$ are the conductive and
convective fluxes, and the dots leave room for any other energy
transport, e.g., by waves and pulsations. The above equation clearly
shows that the sum of all transported energies remains constant, 
since there are no energy sources.

\paragraph{Radiative Atmospheres} In most stellar atmospheres, the
conductive fluxes can be neglected compared to the others, with the
notable exception of specific environments where the conductivity is
large, e.g., coronae and White Dwarfs (degenerate material). Moreover,
velocity fields do not play a decisive role for the energy equation,
as long as $v/c \ll 1$. We first concentrate on those atmospheres
where the energy is transported, to the largest part, by radiation,
i.e., on stars roughly more massive than 1.3 \msun. When integrating
the RT equation (Eq.~\ref{eq_RT}) over solid angle and frequency, one
obtains the energy equation for the radiation field (again adopting $v
\ll c$), which for stationary conditions relates the radiative flux
with the radiative cooling and heating rates, $\Lambda$ and
$\mathcal{H}$ (first equation below). For the considered case of
purely radiative energy transport, flux conservation 
(Eq.~\ref{eq_energy_stat}) then implies
\beq
\label{eq_radeq}
\vec{\nabla} \cdot \mathcal{F}  = 
\int_0^\infty \dd \nu \oint \,\dd \Omega (\eta_\nu - \chi_\nu I_\nu)
= \Lambda - \mathcal{H} \stackrel{!}{=} 0
\qquad \rightarrow L = 4 \pi r^2 \mathcal{F}(r) = 
4 \pi \Rstar^2 \mathcal{F}(\Rstar) = 
4 \pi \Rstar^2 \sigma_{\rm B} \Teff^4 = \text{const}.
\eeq
Here we have defined the stellar luminosity, $L$ (with units
Js$^{-1}$) and the effective temperature, \Teff, the latter as that
temperature where the frequency integrated Planck-function (times
$\pi$, to account for the corresponding flux) equals the photospheric
flux $\mathcal{F}(\Rstar)$ (with $\sigma_{\rm B}$ the Stefan-Boltzmann
constant). Both the first ($\vec{\nabla} \cdot \mathcal{F}$) and the
alternative formulation ($\Lambda - \mathcal{H}$) in
Eq.~\ref{eq_radeq} can be used to derive the atmospheric temperature
structure in an iterative way, until the required condition (= 0) is
fulfilled. Flux conservation is usually utilized at large optical
depths, whereas the second formulation, ``radiative equilibrium''
(considered at low optical depths), states that the absorbed energy
must equal the emitted one. For first estimates and a basic
understanding, one might assume opacities and emissivities to be
frequency independent (``grey''), and couple the frequency-integrated
equation of RT, of its first moment (see textbooks), and of radiative
equilibrium together with some minor approximations to obtain
\beq
\label{eq_greytemp}
T^4(\tau_{\rm grey}) \approx \frac{3}{4}\Teff^4(\tau_{\rm grey} + 
\frac{2}{3}) \qquad
\text{with } \tau_{\rm grey} = \int_z^\infty \chi_{\rm grey} \dd z. 
\eeq
By introducing the widely used \textit{Rosseland mean opacity}, defined
in such a way as to replace the above grey approximation by a more
physical one based on the diffusive character of the radiation field
at large optical depths (for details, see any textbook)
\beq
\label{eq_Ross}
\frac{1}{\bar \chi_{\rm R}} = \Bigl[\int_0^\infty \frac{1}{\chi_\nu}
\frac{\partial B_\nu}{\partial T} \dd \nu \Bigr] / 
\Bigl[\frac{4 \sigma_{\rm B}}{\pi} T^3\Bigr] 
\qquad \rightarrow \qquad T^4(\tau_{\rm R}) \approx
\frac{3}{4}\Teff^4(\tau_{\rm R} + \frac{2}{3}),
\eeq
the temperature structure is now expressed in terms of a meaningful
optical-depth scale. By construction, the Rosseland mean can be
calculated without any radiative transfer, and the harmonic weighting
of the individual contributions accounts for the fact that the maximum flux is
transported in those frequency regions where $\chi_\nu$ is small.
%
%fig_comptstruct
\begin{figure}[t]
\centering
\includegraphics[width=0.6\textwidth,angle=0]{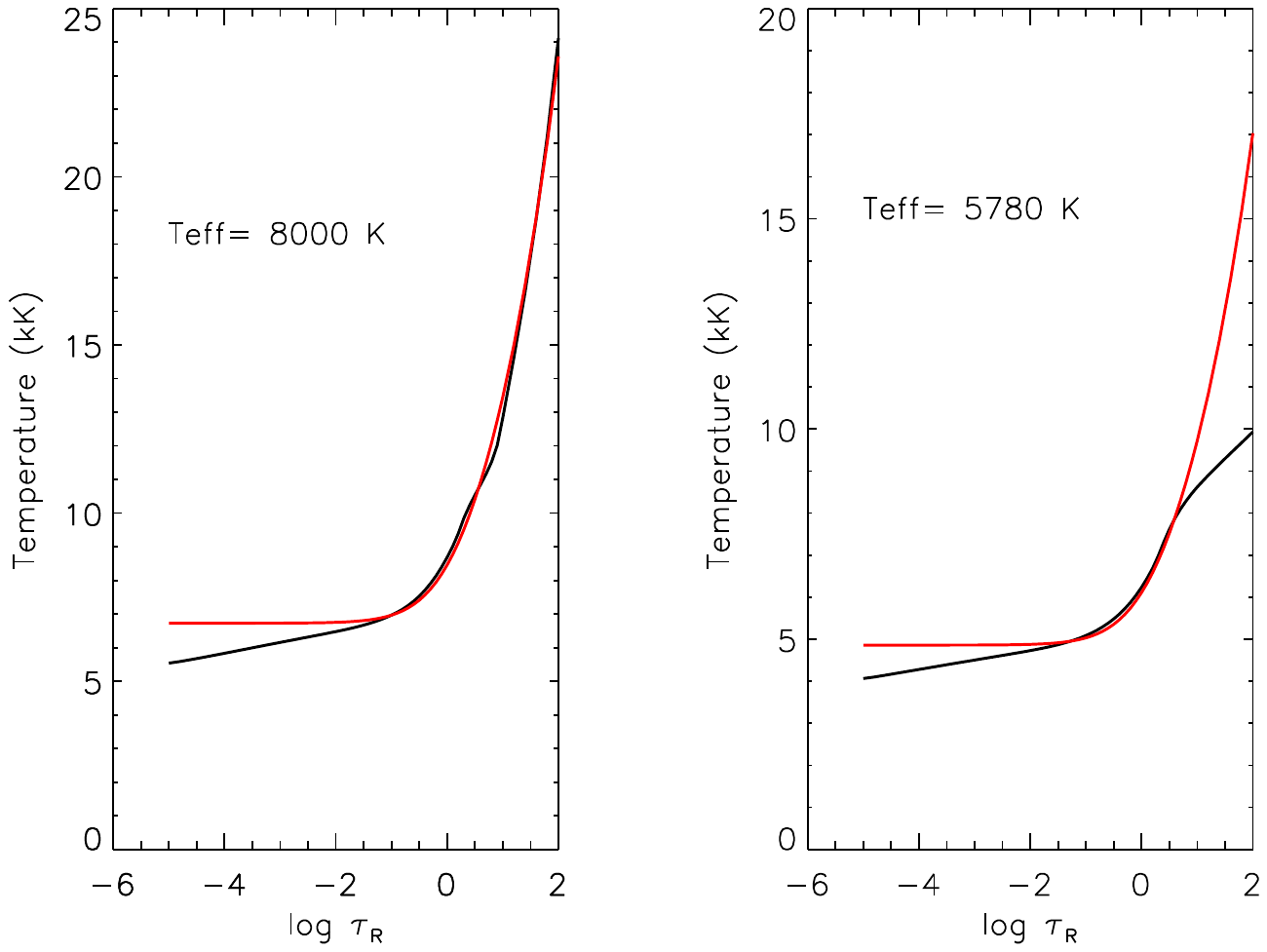}
\caption
{Temperature stratifications as a function of $\tau_{\rm R}$, for an
A-type dwarf model (left) at \Teff\ = 8000~K and \logg\ = 4.0, and
for a solar-type model (\Teff\ = 5780~K and \logg\ = 4.44), right. The
approximate stratification Eq.~\ref{eq_Ross} (right) for radiative
envelopes is displayed in red, and the ``exact'' stratification from a
plane-parallel, hydrostatic MARCS atmosphere (see Sect.~\ref{codes})
in black. Whilst the appproximate radiative stratification matches the
exact one quite well for the A-star model, large differences at high
optical depths are visible for the solar-type model, because of a
convective stratification in the lower photosphere (and below).}
\label{comptstruct}
\end{figure}
%
%fig_solar_ttau
\begin{figure}[t]
\centering
\includegraphics[width=0.6\textwidth,angle=0]{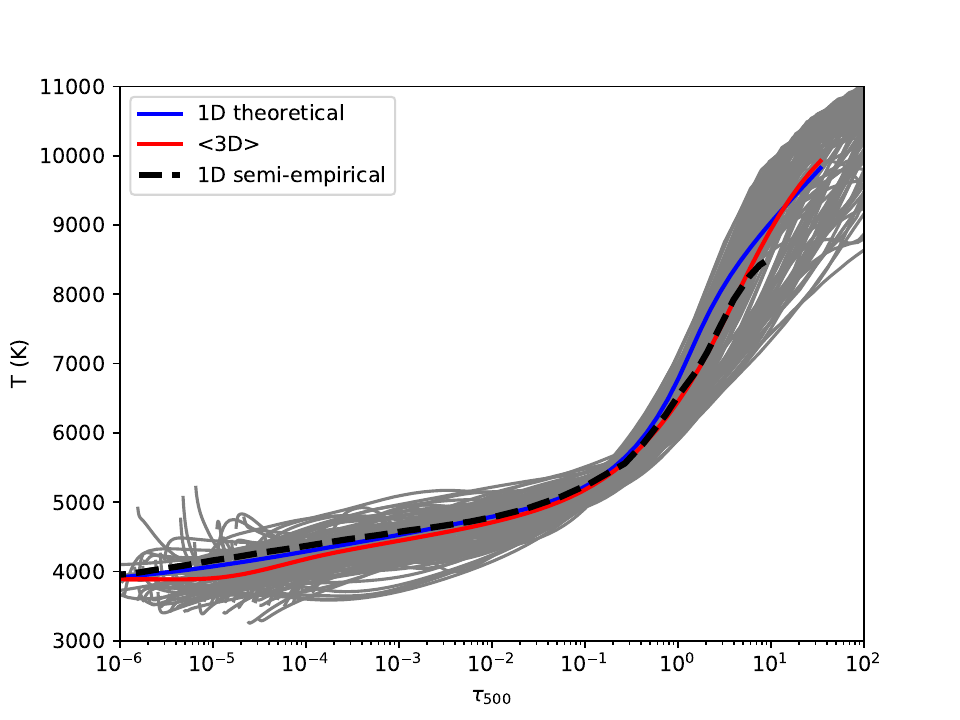} 
\caption
{Temperature stratifications as a function of $\tau_{\rm 500 nm}$ for
the Sun as given by a 1D theoretical model from Kurucz 
\citep[][blue]{2012AJ....144..120M}, the semi-empirical Holweger-M\"uller
model \citep[][black-dashed]{1974SoPh...39...19H}, spatially resolved
profiles from a snapshot of a 3D hydrodynamical simulation
\citep[][grey]{2011SoPh..268..255C} and their spatial average (red).}
\label{solarttau}
\end{figure}

The above temperature stratification is close to being exact at large
optical depth ($\tau_{\rm R} \ga 3{\ldots}5$), but predicts all major
trends also at lower values (see Fig.~\ref{comptstruct}, left panel)
\footnote{We repeat that for an ``exact'' solution, iterative methods
based on Eq.~\ref{eq_radeq}\,(left) need to be employed}. In
particular, the temperature becomes equal to \Teff\, at $\tau_{\rm R}
= 2/3$, which is sometimes used, via Eq.~\ref{eq_radeq}, to
define the stellar radius, namely $\Rstar = r(\tau_{\rm R} = 2/3)$. A
generalization of these results towards spherically extended
atmospheres has been given by \citet{Lucy1971}.

\paragraph{Convective Envelopes} Under conditions typically valid for
low-mass star envelopes, convection becomes the dominant type of
energy transport. When radiative energy transport is the only
competing process, we find (adopting stationary, static and
plane-parallel conditions)
\beq
\label{eq_convtransport}
\vec{\nabla} \cdot [\mathcal{F}^{\rm rad} + \vec{F}^{\rm conv}] = 0
\qquad \Rightarrow \qquad
\frac{\dd F^{\rm conv}}{\dd z} = -\frac{\mathcal{F}^{\rm rad}}{\dd z} =
- \int_0^\infty \dd \nu \oint \,\dd \Omega (\eta_\nu - \chi_\nu I_\nu) = 
- 4\pi \int_0^\infty \dd \nu \chi_\nu (S_\nu - J_\nu) = \mathcal{H} -
\Lambda,
\eeq
i.e., an imbalance between radiative heating and cooling. In a simple
picture, the heat content of lower regions is transported by
``bubbles'' towards outer regions, where these bubbles dissolve and
release their excess energy. The driving force for the outward
propagation of these bubbles is buoyancy, as long as the bubble
has a lower density than the ambient medium. Under the
assumption of pressure equilibrium (requiring slow -- compared to the
speed of sound -- convective velocities) and no energy exchange with
the ambient medium (adiabatic change of state inside the bubbles), one
can formulate a (simplified) criterion for convection to take place,
the so-called Schwarzschild criterion\footnote{Sometimes, the
so-called Ledoux criterion is used, that also takes the effect of
molecular weight gradients into account.}
\beq 
\label{eq_Schwarzschildt} 
\nabla_{\rm a} = \frac{\dd \ln T_{\rm a}}{\dd \ln p} > 1 -
\frac{1}{\Gamma_{\rm ad}} 
= \nabla_{\rm ad}. 
\eeq 
Here, the Nablas are the conventional thermodynamic derivatives, ``a''
means ambient, ``ad'' means adiabatic (for the bubbles), and
$\Gamma_{\rm ad}$ is the adiabatic exponent. 
If this condition is not fulfilled, the stratification becomes
radiative, and $\nabla_{\rm a} \rightarrow \nabla_{\rm rad}$, where
the latter can be estimated from the pressure scale height $H$
(Eq.~\ref{eq_scaleheight}) and the temperature stratification
(approximated in Eq.~\ref{eq_Ross}). For massive and other hot stars, hydrogen (as the
major constituent) is completely ionized, and $\nabla_{\rm ad} \approx
0.4$ turns out to be larger than $\nabla_{\rm rad}$ throughout the 
complete atmosphere. Consequently, hot star atmospheres are
convectively stable, at least under typical conditions (but see
Sect.~\ref{inhomo}). On the other
hand, in the outer photospheres of cool stars, hydrogen is neutral,
but becomes ionized at larger depths. In this case, $\nabla_{\rm ad}$
decreases, with a minimum of 0.07 when 50\% of hydrogen is ionized.
For the Sun, this occurs around 9,000~K, and $\nabla_{\rm ad} <
\nabla_{\rm rad}$, indicating convection is active.
Ionization changes of elements other than hydrogen may also
contribute to the appearance of convective zones.

If an atmosphere turns out to be convectively unstable, convective
transport needs to be included in the modeling. This can be done
either in an approximate way, or by expensive 3-D
radiation-hydrodynamic simulations. The first possibility (still
applied in many numerical codes, both for atmospheric and for stellar
structure/evolution models) bases on the so-called mixing length
theory (MLT), first suggested by \cite{Prandtl1925}.  Within this
approximation, the various gradients are calculated in an iterative
way (e.g., H\&M), from the condition of total flux (radiative +
convective) conservation, accounting for radiative losses during the
lifetime of a mass element.  For low radiative losses, $\nabla_{\rm a}
\rightarrow \nabla_{\rm ad}$ (``efficient convection''), and for
(very) large losses, $\nabla_{\rm a} \rightarrow \nabla_{\rm rad}$ 
(``inefficient convection''). Fig.~\ref{comptstruct} (right panel)
shows a comparison between the temperature structure for a solar-type
atmosphere with (black) and without (red) convection.
  
Multi-D radiation-hydrodynamic simulations allow for a (much more)
precise treatment, but can be performed for only few, representative
frequency bins,
due to the extremely time consuming multi-D RT to calculate the
radiative heating and cooling terms 
($\mathcal{H}-\Lambda \ne 0$ for convection). 
These approaches were pioneered in the late 1980's
\citep{1989ApJ...342L..95S,1991RvMA....4...43S}, and a prominent
application is the comparison with observational data from the Sun
(limb-darkening, line-profile shifts/variations, granulation
patterns), and the re-determination of the solar abundance pattern
\citep{Asplund2009, 2011SoPh..268..255C}. 

Figure \ref{solarttau} illustrates the thermal stratification of a
plane-parallel 1D model atmosphere for the Sun, a 
semi-empirical 1D model, and the average thermal structure from a
snapshot of a hydrodynamical solar simulation. The grey curves
correspond to spatially-resolved stratifications across the 3D
snapshot, and illustrate the horizontal inhomogeneities 
of the simulation.
Hydrodynamical models of cool atmospheres remove the need for 
parameters such as micro and macro-turbulence (see
Sect.~\ref{inhomo}), and predict line profiles that are usually
blueshifted and asymmetric, much like the observed ones. As for the
Sun, thermal inhomogeneities lead also in many other cases to changes
in the inferred elemental abundances and isotopic ratios. 
Various grids of hydrodynamical model atmospheres
for cool stars covering typical parameter ranges with different
metallicities are available (see, e.g.,
\citealt{2009MmSAI..80..711L, 2013A&A...557A..26M, 2024arXiv240507872R}).
\paragraph{Cromospheres and Coronae}
The outermost layers of a low-mass star, sometimes called 
the {\it upper atmosphere}, can be split into the {\it chromosphere} and the
{\it corona}, with an area in between known as the {\it transition 
region}. In these layers the pattern of decreasing temperature seen in
the underlying photosphere is reversed and, for example, in a star
like the Sun, the average temperature increases from a few thousand
Kelvin at the top of the photosphere to millions of Kelvin in the
corona, accompanied by a continuosly decreasing density. In this
tenous environment, photons can travel far from where they have been
emitted before they are absorbed, greatly enhancing departures from
local thermodynamical equilibrium (Sect.~\ref{occnum}), and magnetic
fields play a prominent role. The optical and near-infrared spectra of
cool stars are mostly formed in the photosphere, and hints of the
existence of the upper atmosphere are typically only apparent in the
cores of the strongest lines, such as H$_\alpha$ or the \CaII\ H and K
lines. On the contrary, the upper atmosphere starts to manifest itself
in the near UV, and dominates the solar spectrum in the vacuum UV, 
whereas the corona shows emission lines of extremely ionized metals 
(e.g., \FeXVII). The detailed heating
mechanism (likely related to the dissipation of acoustic and
magneto-hydrodynamic waves) is still subject of current research, but
we note here that because of the low densities, the corona does not
need to be included into a photospheric modeling (though there is
a certain impact from the cromosphere).

\subsection{Opacities and Emissivities}
\label{opa_eta}
The most important quantities required to perform radiative transfer
are opacities and emissivities. Basically, these are the sums of all
individual contributions at a given frequency, and can be calculated
from the cross-sections (atomic/molecular physics) times the
occupation numbers (number densities of atoms and molecules 
populating each of the absorbing and emitting levels). For the line
case, (normalized) profile functions are required as well, providing
the probabilities for an absorption/emission event at frequencies away
from the line center. In addition, scattering processes need to be
accounted for.

\subsubsection{Line Processes}
\label{lines}
For transitions between bound levels, the individual line opacities and
emissivities can be calculated from 
\beq 
\label{eq_lineopa}
\chi_\nu^{\rm line} = \frac{h\nu_{\rm ul}}{4\pi}\phi(\nu) 
\Bigl[n_{\rm l}B_{\rm lu}-n_{\rm u}B_{\rm ul}\frac{\psi(\nu)}{\phi(\nu}\Bigr], 
\qquad
\eta_\nu^{\rm line} = \frac{h\nu_{\rm ul}}{4\pi}\psi(\nu) n_{\rm u}A_{\rm ul},
\eeq 
where $\nu_{\rm ul}$ is the frequency at line center, $\phi$ and
$\psi$ are the absorption and emission profile functions, $n$ the
occupation number densities for the lower (``l'') and upper (``u'')
levels of the considered transition, and $B_{\rm lu}, B_{\rm ul},
A_{\rm ul}$ the \textit{Einstein coefficients} (atomic properties
closely related to the transition probability, and independent of the
thermodynamic state) for absorption, induced and spontaneous emission,
respectively. Induced emission is included here as negative
absorption. Under typical conditions in stellar atmospheres (no
correlation between absorbed and emitted frequencies, so-called 
\textit{complete redistribution}, see textbooks), $\psi \rightarrow
\phi$. From simple arguments, it is easy to show that there are two
relations connecting the three Einstein coefficients, and only one
needs to be known for the line opacity/emissivity; the line source
function does not depend on any of those at all. $B_{\rm lu}$ can be
also expressed in terms of the classical cross-section from
electrodynamics and the quantum-mechanical oscillator-strength, $f$. 
The profile function, $\phi(\nu)$, is a convolution of different
broadening functions, in particular natural line broadening
(basically, a quantum electrodynamic effect, heuristically resulting
from the finite life-time of the emitting state and Heisenberg's
uncertainty principle) and various kinds of collisional broadening
(perturbation of radiating particles by other plasma components, e.g.,
Stark, resonance and van der Vaals broadening). Except for linear
Stark-broadening, all of these can be described by a Lorentzian
(Cauchy) distribution. Most important for the line cores is 
Doppler-broadening, which accounts for the Doppler effect due to the
thermal (Maxwellian) velocities of the absorbing and emitting
particles. The convolution of the latter Gaussian with the Lorentzian 
from a above is called a \textit{Hjerting} or \textit{Voigt profile}.
 
When the gas reaches temperatures cool enough (typically about
4000~K), molecules become progressively important. Molecular
electronic transitions split in a myriad of vibrational and rotational
components that form absorption bands that can span a broad spectral
range, impacting significantly the emerging SED of the atmosphere and
the energy balance. The chemical equilibrium among all the relevant
molecular species, which are many, needs to be computed in detail, and
the outcome affects atomic absorption lines, since atoms trapped in
molecules no longer contribute to atomic transitions. Fortunately,
quantum mechanical calculations have improved their accuracy in recent
years, and are now providing detailed line lists for the most
important molecular species (see,e. g.,
\citealt{2018Atoms...6...26T}).

\subsubsection{Continuum Processes and Scattering}
Bound-free and free-free opacities/emissivities can be calculated from
the \textit{Einstein-Milne} relations, and also involve products of
occupation numbers and quantum-mechanical cross-sections. Bound-free 
cross-sections are typically different from zero for frequencies
larger than the corresponding ionization threshold (but can extend
also to lower frequencies, because of so-called resonances), and
free-free processes increase quadratically with density and
wavelength, dominating the IR and radio domains. In cool stars, the
free-free and bound-free opacities from the H$^-$ ion, with one bound
state at 0.75~eV, have to be accounted for (essential for solar-type
stars in the optical and near IR).

A variety of potential scattering processes additionally contribute to
the total opacity and emissivity. Most important is the scattering by
free electrons, which in the range roughly below 12.4~keV becomes
coherent and is called Thomson scattering. The corresponding
scattering opacity, $\sigma_\nu$, only depends on cross-section and
electron density (low in cool star atmospheres), and the emissivity
can be fairly well approximated in terms of the mean intensity,
\beq
\chi_\nu^{\rm Th} = \sigma_{\rm Th} n_{\rm e} \qquad \text{with}
\qquad \sigma_{\rm Th} = \frac{8\pi}{3} r_0^2 \approx 6.65\cdot 10^{-29} {\rm m}^2,
\qquad
\eta_\nu^{\rm Th} \approx \chi_\nu^{\rm Th} J_\nu
\eeq
($r_0$ is the classical electron radius). Other scattering processes
are the high-energy generalizations of Thomson-scattering (Compton-
and Klein-Nishina scattering), Rayleigh scattering (basically
atomic/molecular line absorption/emission far from their central 
frequency), and sometimes also Raman-scattering (inelastic
scattering by molecules). After summing up the individual
contributions from lines, continuua and scattering processes, the
total source function does not only depend on atomic quantities and
occupation numbers, but also on the mean intensity (because of the
scattering terms), which immediately shows the necessity of an
iterative procedure.

\subsection{Occupation Numbers}
\label{occnum} 
As long as a plasma is dominated by collisions at a given point, i.e.,
if there are (in each transition!) much more collisional than
radiative processes, the plasma will be in \textit{local thermodynamic
equilibrium} (LTE). In this case, all occupation numbers (ionization
and excitation) can be calculated from the well-known Saha-Boltzmann
equation, in dependence of local electron density and temperature.
If the above condition is no longer fulfilled for all
transitions, either part or even all atomic/molecular levels require a
so-called non-LTE (NLTE) or \textit{kinetic equilibrium} treatment. As
a rule of thumb, atmospheric LTE conditions (high densities and low
temperatures) are mostly met in dwarfs and partly giants of spectral
type late B and cooler (for Galactic metallicities). For the rest,
NLTE is required. We note that even for relatively cool stars such as
the Sun, detailed calculations including 3-D modeling show a better
reproduction of specific lines (e.g., strong \FeI, \CaI\ and \CaII\
lines) when calculated in NLTE (e.g., \citealt{Bergemann2012}). We
also note that in the deepest part of stellar atmospheres (lower
boundary), LTE prevails, since the density increases more rapidly than
temperature.  

Under NLTE conditions, the calculation of occupation numbers and
radiation field has to proceed (iteratively) in parallel (together
with temperature and density stratification): Saha-Boltzmann is no
longer valid, and the occupation numbers have to be derived from the
so-called \textit{rate equations}. Under stationary conditions, i.e., as
long as kinematic time-scales are much larger than atomic
ones\footnote{such conditions are no longer met in SN remnants, e.g.,
\citet{Hillier2012}}, these are given by
\beq
\label{eq_rate}
\sum_{j \ne i} n_{\rm i} P_{\rm ij} = \sum_{j \ne i} n_{\rm j} P_{\rm
ji} \qquad \forall \text{levels i}. 
\eeq
Here, the indices (i,j) refer to all levels considered in the model
atom (a theoretical description of levels and potential transitions,
see Sect.~\ref{atomdat}), $n_{\rm i}$ are the corresponding occupation
number densities, and $P_{\rm ij}$ the transition rates from level i
to level j, for radiative and collisional bound-bound and bound-free
processes.  Eq.~\ref{eq_rate} states that the number of all possible
transitions from level i to other levels j needs to be balanced by the
number of transitions from all other levels j into level i. In
mathematical terms, this is a linear equation system, which needs to
be closed with an equation of particle (or charge) conservation

There are at least three major challenges when solving
Eq.~\ref{eq_rate}: (i) For each(!) transition (and usually $10^5
{\ldots} 10^6$ transitions need to be considered), corresponding
atomic data (radiative and collisional cross sections) need to be
present. (ii) All radiative rates depend on the mean intensity, and
the RT needs to be solved for a large frequency range, with high
resolution to resolve all line cores and ionization edges
(computational time!). (iii) Since the occupation numbers depend on
the transition rates and thus the radiation field, and the radiation
field depends on the opacities and thus occupation numbers, a clever
iteration scheme\footnote{either by means of the so-called Accelerated
Lambda-Iteration \citep{Werner1985}, well-known in the realm of
elliptic partial equations under the name ``Jacobi iteration'', or in
terms of the so-called complete linearization method, firstly
suggested by \citet{Auer1969}} needs to be established.
NLTE calculations for molecular species were traditionally considered 
as out of reach due to the vast numbers of levels and transitions involved.
However, there are brave efforts on this front giving promising results 
\citep{2003ASPC..288..339S, 2016SoSyR..50..316O, 2023A&A...670A..25P}.

\section{Advanced Topics} 
\label{advanced} 
\subsection{Stellar Winds}
\label{winds}
Stellar winds (= mass outflows) are ubiquitious throughout the
Hertzsprung-Russel diagram, and indeed it is difficult (if not
impossible) to find a star that has no wind at all. Such winds can
have vastly different mass-loss rates and terminal velocities, but if
the wind-strength is not very low, they have to be included (at least
in principle) into atmospheric calculations. Using the 1-D equation of
motion together with the equation of continuity
(Eqs.~\ref{eq_motion}ii/\ref{eq_motion}i) and the equation of state,
$p=\rho v_{\rm s}^2$, an alternative formulation reads
\beq
\label{eq_motion1}
\Bigl(1-\frac{v_{\rm s}^2}{v^2}\Bigl)\,\rho v(r) \frac{\dd v}{\dd r} = 
-\frac{GM}{r^2} + g^{\rm o}(r) + \frac{2 v_{\rm s}^2}{r} -
\frac{\dd v_{\rm s}^2}{\dd r}
\eeq
where we have split the external accelerations into gravity and other
accelerations ($g^{\rm o}$), and rewritten the gas pressure gradient. 
Basically, four different solutions are possible, but when
concentrating here on a potential mass outflow, we see that in order
to switch from subsonic to supersonic velocities (lhs), the rhs has
also to switch sign: from the sonic point on, the
sum of the 2nd to 4th term has to become larger than gravity.

{\bf Pressure Driven Winds.} For cool main sequence stars, external
forces besides gravity (such as radiative acceleration) are weak and
usually negligible, and we restrict ourselves to the case of a wind
initiated by a high-temperature corona, as firstly realized by
\citet{Parker1960}. If such a high temperature can be maintained by
heating and thermal conduction (see Sect.~\ref{energy}), in an
isothermal wind the equation changes sign at the sonic (``critical'')
point, where $r_{\rm s}=GM/(2v_{\rm s}^2)$. For an average temperature
of $2\cdot10^6$~K then, $v_{\rm s}$ $\approx$ 165~kms$^{-1}$ and
$r_{\rm s}$ $\approx$ 3.5~\rsun. The low coronal density implies a low
mass-loss rate of roughly $10^{-14}$\msunyr, irrelevant on
evolutionary time-scales of $10^{10}$yr, and the typical terminal
velocity is around 500~kms$^{-1}$. Despite the small
implications for the Sun's evolution, the solar wind has an obvious impact on
communications and space weather.
 
{\bf Line-Driven Winds.} From improved rocket- and first
satellite-based UV observations in the 1960s and 70s on, it became
clear that massive OB-stars possess quite strong winds, because of
their strong UV P~Cygni profiles\footnote{broad, Doppler-shifted
absorption bluewards from the rest-frame transition frequency, 
overlayed by emission extending to the red}. As they usually lack a
hot corona, other acceleration mechanisms need to be invoked.
Pioneering work to explain these winds by radiative line-driving was
performed by \citet{Lucy1970} and \citet{CAK1975}, and although 
various improvements to the original formulations have been included
meanwhile, the basic theory still holds, at least qualitatively.
Because of the large oscillator-strengths of line transitions compared
to continuum cross-sections, and because of the high luminosity of
massive stars ($10^4$ to few $10^6 \lsun$), the radiative line
acceleration (momentum transfered by line-scattering and absorption)
is sufficient to accelerate winds to high speeds (couple of 100 to
3000 kms$^{-1}$) at significant mass-loss rates ($10^{-8}$ to
$10^{-5}$ \msunyr), which is of major concern for their evolution. For
details, we refer to \citet{Puls2008}, \citet{Vink2022}, and the chapter about
``Stellar Winds'' within this Encyclopedia of Astrophysics. We only note
here that (i) mostly metal ions are accelerated, which transfer their
momentum to the bulk matter (H and He) via Coulomb collisions, and
that (ii) without the distance-dependent Doppler shift between the
stellar rest frame and the moving ions line acceleration would not
work, since then the lines would become already saturated in the upper
photosphere, and could no longer accelerate the wind at larger
distances from the star. Because of their large densities, line-driven
winds need to be accounted for when modeling the atmospheres of hot
massive stars (see Sect.~\ref{hydro}), at least when the actual
mass-loss rate is larger than few times $10^{-8} \msunyr$.
For lower mass-loss rates, most diagnostic line-features form in the
quasi-hydrostatic part of the atmosphere, and a plane-parallel,
hydrostatic modeling becomes possible \citep{Puls2009}. 

{\bf Cool Star Winds.} 
In Red Supergiants and Asymptotic Giant Branch (AGB) stars, their
extreme radii and very low gravities enable the formation of giant
convective cells (with $H/R \approx 0.2$), and 3-D simulations
\citep{Freytag2023} show that these can lift material to cool regions
($T \la $1,500~K) where dust can form. Because of high luminosities,
a wind is accelerated beyond escape by radiative dust-driving,
where the terminal velocities are low (few tens of kms$^{-1})$ but the
mass-loss rates high ($10^{-7} {\ldots} 10^{-5}$\msunyr). For AGB
stars, a ``pre-acceleration'' by their slow pulsations steepening into
strong shocks is additionally required, and in Red Supergiants an
alternative acceleration might be due to turbulent pressure
\citep{Kee2021}. Winds from ``normal'' Red Giants, on the other hand,
are not understood until now. 

{\bf Continuum Driven Super-Eddington Winds.} All above
radiation-driven winds require the presence of metals, and are
otherwise very weak or even absent. An alternative wind-mechanism in
hot massive stars which does not involve any metals -- and thus might
be particularly important for the very first stellar generation -- is
(theoretically) given by an acceleration due to Thomson-scattering. If
a hot massive star is near or above the \textit{Eddington
limit}\footnote{$g_{\rm rad}^{\rm Th}/g_{\rm grav} = 1$; for ionized H
and He, this condition is independent of $r$, since both quantities
dilute with $r^{-2}$}, an outwards decreasing porosity
(Sect.~\ref{inhomo}) can lead to a Thomson acceleration that is below
gravity in photospheric regions, and above beyond, such that a wind
can form (see Eq.~\ref{eq_motion1}), instead of an expanding envelope.
For details and implications, we refer to \citet{Owocki2004} and
\citet{Smith2006}.

\subsection{Impact of Rotation and Magnetic Fields}
\label{rot}
Until here, we have built upon a rather simplified atmospheric model,
neglecting rotation, magnetic fields, and (see Sect.~\ref{inhomo}) 
inhomogeneities (e.g., turbulence). Magnetic fields of considerable
strength are rare in hot massive stars (detected for $\sim$10\% of
main-sequence objects, e.g., \citealt{Grunhut2017}), and their origin
is still debated \citep[e.g.][]{Schneider2016}, whereas they play a
very important role in cooler low-mass stars (like the Sun), affecting
the whole atmosphere and being responsible for the presence of
sunspots and coronae. Rotation, on the other hand, is slow in low mass
stars, but can reach hundreds of \kms in massive hot stars, affecting
their geometrical shapes and structure.

In 1-D models, magnetic fields (if sizeable) might be accounted for by
means of the corresponding magnetic pressure (adopting a simplified
field geometry). Magnetic fields affect the line-profiles by Zeeman-splitting
(increasing with field-strength and square of transition wavelength),
but, most importantly, lead to circular polarization, which can be
used to infer magnetic field strengths averaged over the stellar disk
(e.g., \citealt{Wade2016}). 

1-D rotation is usually considered in an implicit way only: the
atmospheric parameter $g_{\rm grav}$ is re-interpreted as an effective
gravity, $g_{\rm eff} = g_{\rm grav} - g_{\rm cent}$, to be derived
from quantitative spectroscopy (see Sect.~\ref{methods}). The actual
$g_{\rm grav}$ is then found by correcting $g_{\rm eff}$ with an
estimate for the centrifugal acceleration $g_{\rm cent}$, averaged
over the apparent stellar disk, and in dependence of $v_{\rm rot} \sin
i$ (\citealt{Repolust2004}, Appendix A). Here, the latter quantity is
the projected rotation speed at the stellar equator, with $i$ the
angle between observer and stellar rotational axis, and is the only
quantity that can be actually measured from observations (as long as
no additional information, e.g., from the motion of stellar spots, is
available).  To this end, and to allow for a meaningful spectroscopic
analysis, ``non-rotated'' 1-D synthetic spectra need to be convolved
with a corresponding broadening function.  The convolved line-profiles
extend to $\pm v_{\rm rot} \sin i$, and their shapes\footnote{though
often contaminated by other processes, in particular macro-turbulence
(see below) and instrumental resolution} (see Fig.~\ref{profiles},
right panel) allows to measure the projected rotational speed, by
adapting $v_{\rm rot} \sin i$ until the observed widths, shapes and
depths are reproduced. Alternatively, a Fourier method can be used
\citep{SSD2007}. One important aspect of rotational broadening is its
conservation of equivalent width, i.e., the area of the normalized
profile below or above the continuum.

Because of centrifugal acceleration, the stellar surface becomes
deformed, with a maximum ratio $R_{\rm eq}/R_{\rm pole} \approx 1.5$
at \textit{critical rotation}, i.e., when the total acceleration at
the equator becomes zero. Estimates for the surface radius as a
function of rotational speed and latitude can be found, e.g., in
\citet{Cranmer1995} and references therein. Moreover, as already shown
by \citet{vonZeipel1924}, and somewhat generalized by
\citet{Maeder1999}, the surface flux of a rotating star is
proportional to the effective gravity. Since the flux can be expressed
in terms of \Teff, the effective temperature is hotter at the pole
than at the equator. Typically, this \textit{gravity darkening} effect
becomes noticeable for rotational speeds $\ga 70\%$ of the critical
one.\footnote{see \citet{MM2000} for an in-depth study how to
calculate the critical velocity.} Photospheric models aiming at
including these variations create a suitable mesh over the distorted
surface (e.g., by tesselation), and populate this mesh with the
emergent intensities from model atmospheres with parameters $g_{\rm
eff}$ and \Teff\ as a function of position. An integration over the
individual intensities results in the final SED, depending on $\sin i$
\citep[e.g.,][]{Abdul-Masih2020}.

Both rotation and magnetic fields affect the stellar winds from hot
stars. Contrasted to the oblate stellar surface resulting from rapid
rotation, line-driven winds are predicted to become prolate, because
of the larger illumination at the poles (gravity darkening). The polar
mass-loss rate becomes larger than the equatorial one, though the
total mass-loss rate is barely affected, except for rotational
velocities close to critical. Since in line-driven winds terminal
velocities scale with $g_{\rm eff}$ (now including the Thomson
acceleration), also the velocities are predicted to become larger at
the pole compared to equatorial regions. This difference leads to
non-radial terms in the line-acceleration, resulting in a slow
velocity component into the polar direction. For details, we refer to
\citet{Cranmer1995} and \citet{Puls2008}.

In a series of papers, ud-Doula and co-workers studied the impact of
magnetic fields on line-driven winds (starting with
\citealt{udDoula2002}), and showed, as one of their major results, that
magnetic fields with strengths of few $10^{-2}$\,T are able to quench
the mass-loss rates significantly. Moreover, because of magnetic
braking, the rotational speed decreases on evolutionary time-scales,
in dependence of $\dot M$.

\subsection{Inhomogeneities}\label{inhomo}
One last approximation in standard model atmospheres is the assumption
of homogeneity. This assumption concerns at least two different
aspects. On the one hand, conventional approaches adopt the chemical
abundance pattern to be the same everywhere in the atmosphere. For
atmospheres where mixing motions are weak,
however, gravitational settling and radiative levitation might set in,
and the abundance pattern becomes a function of depth
\citep{Michaud1970}. Well-known examples for this process are Ap-stars
or, more generally, Chemically Peculiar stars on the main sequence and
in the subdwarf domain \citep[e.g.,][]{Michaud2011}.

A second aspect of potential inhomogeneities is related to density-
and velocity fluctuations, e.g., turbulence and patterns due to strong
instabilities and potential shock waves. For cool stars, the most
well-known instability is the convective one (see Sect.~\ref{energy}),
which in solar-type stars leads to granulation. Before multi-D
simulations became available (but also still at present
time), the impact of fluctuating or turbulent velocity fields was
``artificially'' included into the calculation of stellar spectra, 
as so-called micro- and macro-turbulence. Micro-turbulence
($v_{\rm mic}$) adopts that the size of turbulent cells is small
compared to the photon mean free path, and is included into the
individual profile functions (used within the RT) by an additional
term, such that the effective thermal velocity is given by $v_{\rm
th}^2 = (2 k_{\rm B} T/m + v_{\rm mic}^2)$, with $m$ the
atomic/molecular mass. By construction, micro-turbulence does not
conserve the equivalent width. Macro-turbulence, $v_{\rm mac}$, on the
other hand, is supposed to be made up by turbulent cells that are
typically larger than the photon mean free path, and is, in parallel with
rotation, treated by a final convolution of line-profiles with a
Gaussian of width $v_{\rm mac}$ (or a corresponding radial-tangential
profile, e.g., \citealt{Gray2021}), thus conserving the equivalent
width. One of the big breakthroughs of cool-star multi-D simulations
was their capability to reproduce, in particular for the Sun, the
observed profile shapes, widths and shifts, without requiring any
artificial non-thermal velocity field, i.e., without introducing any
micro- and macro-turbulence \citep{Asplund2009}.

\begin{figure}[t]
\centering
\begin{minipage}{8cm}
\resizebox{\hsize}{!}
  {\includegraphics[angle=180]{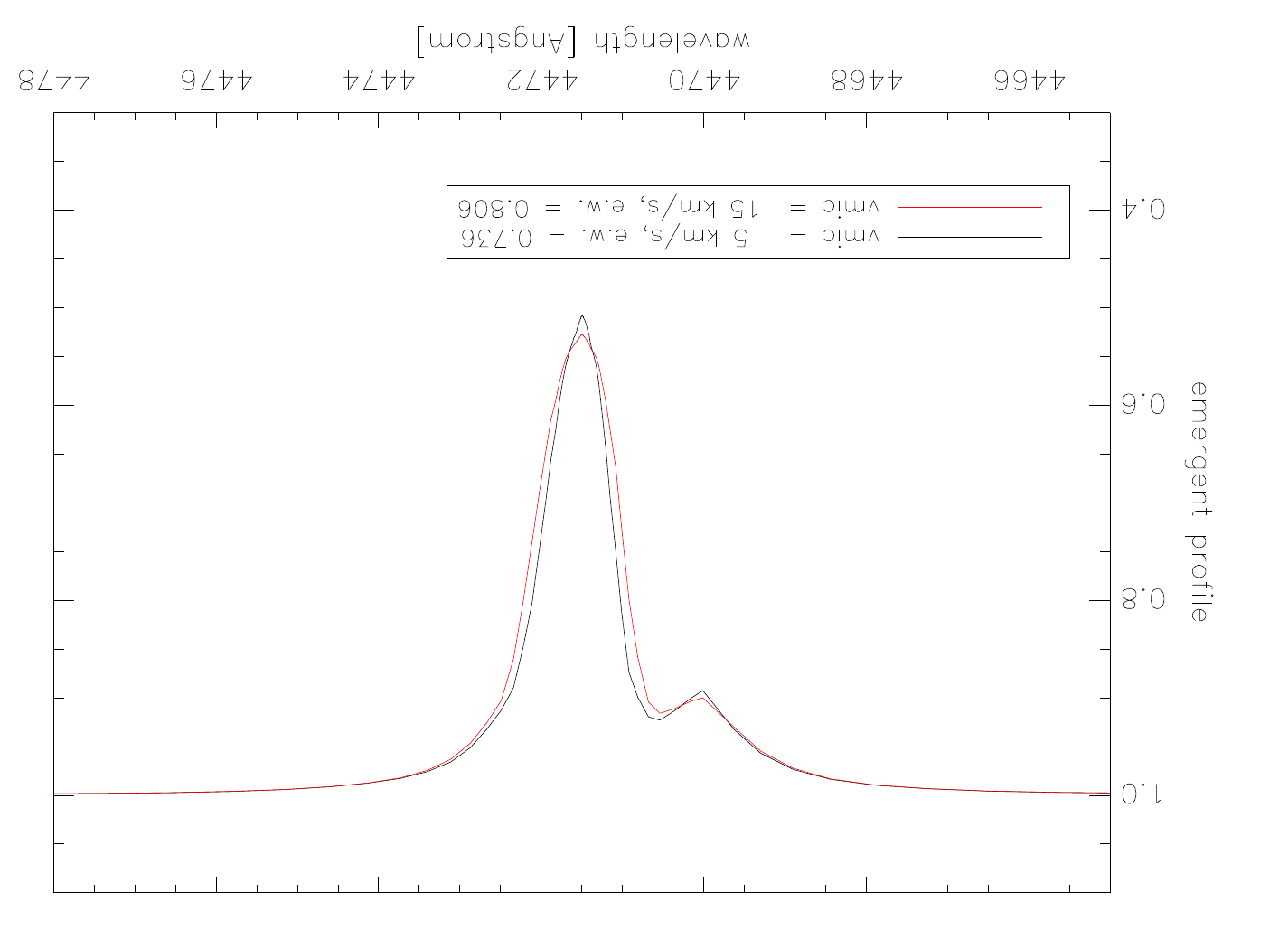}} 
\end{minipage}
\begin{minipage}{8cm}
\resizebox{\hsize}{!}
  {\includegraphics[angle=180]{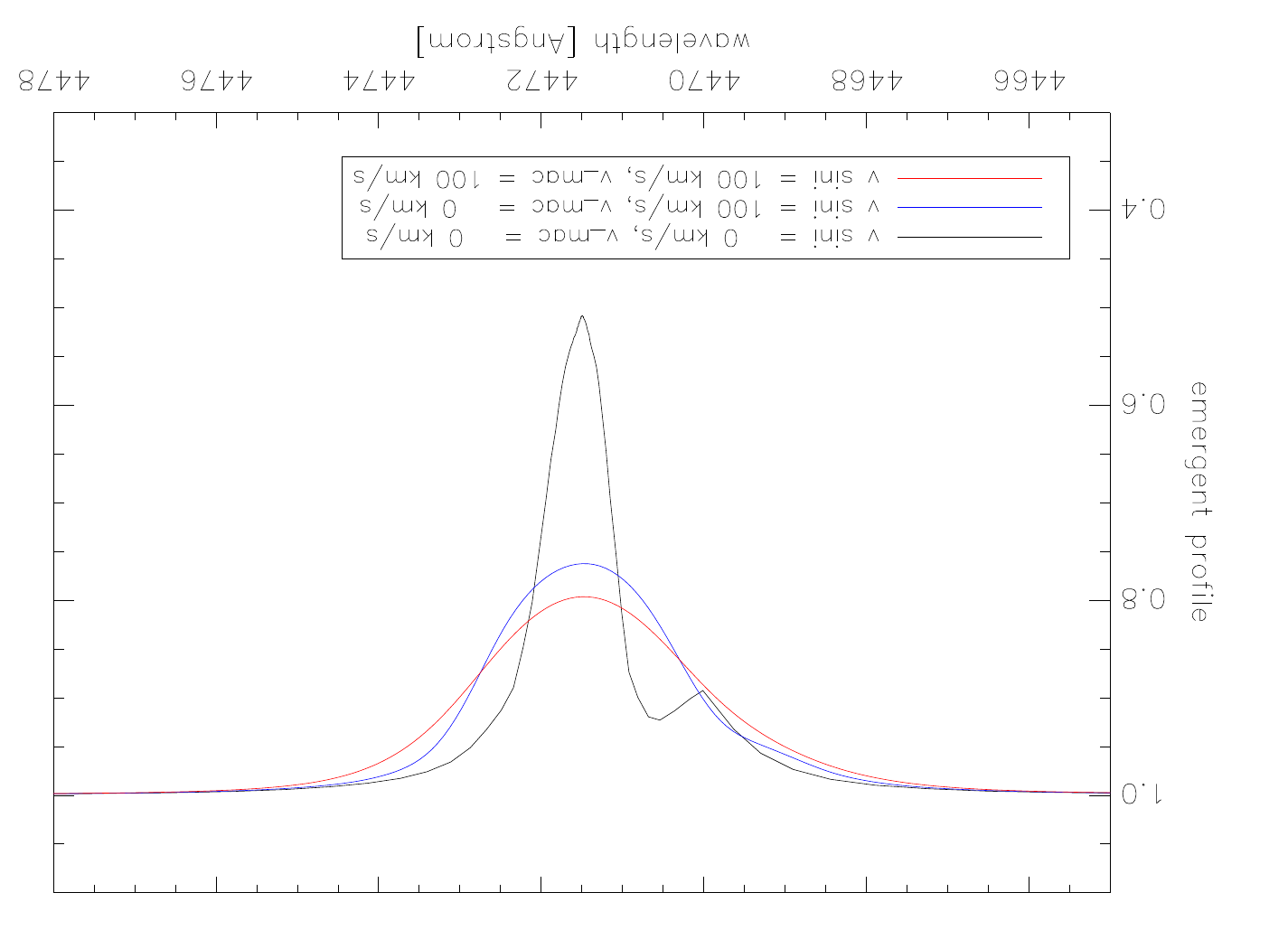}} 
\end{minipage}
\caption{Synthetic line profiles of \HeI\ 4471 $\AA$, for an O-star with
\Teff\ = 35,000~K, \logg\ = 4.0 (cgs), and a wind of
intermediate strength. The second line component to the left of the
main feature is a forbidden one. Left: impact of micro-turbulence,
with e.w. the equivalent width in $\AA$. Right: impact of rotation and
macro-turbulence.  For values and colors, see legend.}
\label{profiles}
\end{figure}

Modelling of hot star atmospheres requires as well the inclusion of
micro- and macro-turbulence (Fig.~\ref{profiles}). First, the derived
$v_{\rm mic}$ values seem to increase with temperature and luminosity
class, from values at or close to zero for late B-type dwarfs to
values of 20~kms$^{-1}$ for OB-supergiants, and there are indications
that micro-turbulence might be related to sub-surface convection zones
\citep{Cantiello2011}. In many stars, the inferred macro-turbulent
velocity is highly supersonic, with typical maximum values of roughly
100 kms$^{-1}$ \citep{SSD2007}. Currently, there are two explanations,
namely either a relation to non-radial pulsations \citep{Aerts2009},
or, again, a relation with sub-surface processes. Indeed, first
pioneering 2-D simulations for hot star atmospheres by
\citet{Debnath2024} including the Fe opacity peak around 200,000~K
resulted in large turbulent velocities in the stellar photosphere,
comparable to the observed values of $v_{\rm mac}$.
%fig_profiles

A major result relevant for hot star atmospheres close to the
Eddington-limit and relevant in the context of continuum-driven
winds (see Sect.~\ref{winds}) was found by \citet{Shaviv1998}. When considering 
a locally inhomogeneous medium, he showed that because of the porous
structure of the medium, the average radiation force becomes less than
in a homogeneous one, such that the mean luminosity can actually exceed the
classical Eddington-limit. 

Inhomogenieties play also an important role in line-driven winds of
hot stars. As shown by \citet{ORI1984}, the driving agent,
line-acceleration, is prone to a strong instability, the so-called
\textit{line-deshadowing instability}: Perturbations in velocity (and
velocity-gradients) lead to perturbations in the line-acceleration,
leading to further growth. The perturbed medium steepens into reverse
shocks, which after (mostly radiative) cooling give rise to
over-densities, compared to the mean flow. The predicted
shock-temperatures are around a couple of million Kelvin, and the
implied X-ray emission has been actually observed in all massive
OB-stars. In a simple picture, the flow then consists of a collection
of dense clumps (with velocities slightly below the smooth-wind
predictions), and a fairly thin and fast inter-clump material. First
1-D hydrodynamic simulations have been performed by \citet{OCR1988}
and \citet{Feldmeier1995}, and meanwhile few multi-D simulations have
been undertaken and studied as well (e.g., \citealt{Sundqvist2018}).

This conceptually different, inhomogeneous wind structure needs to be
included into the \textit{unified} (photosphere + wind) atmospheric
models of hot massive stars, since in such an inhomogeneous (and
partly shocked) medium ionization and excitation are significantly
different compared to the homogeneous case. This inclusion is
performed by means of a simplified description in terms of
over-densities, density contrasts, and volume filling factors (both
for the clumps and for the shocks), where most atmospheric codes rely
on the assumption of optically thin clumps. However, also the effects
of clumps becoming optically thick (porous) at specific wavelengths
(particularly in UV spectral lines) can be treated in an approximate
way. For details, we refer to \citet{Puls2008} and references therein,
and to \citet{Oskinova2007}, \citet{SP2018}, and \citet{Brands2022}.
We only mention here that the spectroscopically inferred mass-loss
rates using such clumped models are lower (typically by a factor two
to three) than derived from smooth-wind models, and that the
interclump density seems to be larger than previously expected (e.g.,
\citealt{Hawcroft2021}). We stress that uncertainties regarding the
inhomogeneous structure of massive star winds lead to sizeable
uncertainties in the derived mass-loss rates, with partly drastic
effect when transferred to evolutionary calculations.

\subsection{Binarity/Multiplicity}
\label{binarity}
About half of all solar-type stars are binaries (though the majority
is widely separated, e.g., \citealt{Moe2017}), and the binary fraction
of massive stars is even higher: 70\% of the Galactic O-stars are
affected from binary interactions, and 24\% will finally merge
\citep{Sana2012}. If the components are close enough, such
interactions need to be included into the modelling of the
atmosphere(s) and into the spectroscopic analysis. Because of tidal
effects, the surfaces become distorted (in addition to rotational
effects), but now as a function of phase. Moreover, the companion
illuminates the atmosphere from outside, and the corresponding
intensity needs to be specified (again as a function of phase), also
accounting for reflection effects. The basic approach is similar to
the treatment of (fast) rotation, namely by creating a suitable mesh
over the distorted surface, and by populating this mesh with
consistently corrected emergent intensities from appropriate,
position- and phase-dependent models (\citealt{Wilson1971}, {\sc
phoebe}-code by \citealt{Prsa2016}). This procedure needs to be
applied to both components, and the common SED is calculated by
integrating over all visible intensities from the complete system as
seen by an observer at the considered phase
\citep[e.g.][]{Abdul-Masih2020}. If both companions have quite
different spectral types, and we were interested in specific line
profiles that form in only one component, it is often sufficient to
consider only the continuum light from the other (or others, if a
``third light'' was present). The same is true if an accretion disc
was present in the system.

\subsection{Atomic and Molecular Data}\label{atomdat}
Under the assumption of LTE (Sect.~\ref{occnum}), the local radiation
field and the occupation numbers are in balance\footnote{except for
the outermost atmospheric regions where the mean intensity is strongly
affected by photon escape}, collisions play a dominant role, and there
is only one temperature describing the velocity distribution for all
particles: ions, electrons and molecules. The level populations
can then be readily computed using the Saha-Boltzmann equations
(with just ionization and excitation energies, statistical
weights and partition functions), and the strength of the spectral
lines can be determined by solving the RT equation given only the
transition probabilities (oscillator strengths), and the damping
constants for strong lines. 

Elastic collisions shift and broaden the lines. While collisional
shifts have a minor impact, collisional damping is usally important
and leaves an obvious mark in stellar spectra. The most important
perturber in low-mass stars are hydrogen atoms, while in hotter stars
it is mainly free electrons that have a dominant effect. Detailed
quantum-mechanical calculations of the interactions provide in many
cases reliable data, and in some cases approximate treatments are
available.

When LTE does not hold, photons can travel far from where they 
have been emitted and deposit energy. The balance between radiative and
collisionally-induced transition needs to be determined for each point
(rate equations, Sect.~\ref{occnum}), and many more data become
necessary. The full collection of data required to solve the rate
equations for each of the relevant ions involved is known as a {\it
model atom}, and includes the energy levels, photo-ionization
cross-sections (including their frequency dependence), collisional
ionization strengths, radiative transition probabilities, and
collisional excitation strengths. Inelastic collisions with electrons
are dominant in intermediate and hot stars, while in cooler types (F
and later) collisions with hydrogen atoms can be very relevant, in
particular for metal-poor stars. In stars with very low masses, 
molecular perturbers may need to be considered as well. 

Ab initio quantum mechanical calculations can provide reliable 
radiative transition probabilities for light atoms/ions, but for
complex atomic systems laboratory measurements are usually required
for accurate results. Theoretical calculations of atomic structure
nowadays give solid predictions for electron-ion recombination rates
and photoionization cross-sections \citep[see,
e.g.][]{2016NewA...46....1N, 2017A&A...606A..11B}, and these can be
checked against detailed spectrophotometric observations of stars
\citep{2023Atoms..11...61A}. Reliable calculations are now becoming
available also for inelastic collisions with hydrogen atoms (e.g., 
\citealt{2012A&A...541A..80B,2016PhRvA..93d2705B,2018A&A...610A..57B}).
Molecules are included in synthetic spectra assuming LTE for the most
part, but their contribution to the equation of state and to the total
opacity is now modeled in far more detail thanks to efforts to improve
partition functions and detailed line lists, by many but in particular
by the Exomol group \citep{2024arXiv240606347T}.

\section{Quantitative Spectroscopy: Determination of Atmospheric Parameters
and Abundances}\label{stellarparams}
\subsection{Observational Data}
\label{obs}
Quantitative spectroscopy is based on observational data, mostly high
resolution spectra with a high signal to noise ratio.  For many
purposes, the optical and near IR range (if molecular bands shall be
analysed) is sufficient. For the analysis of hot stars (with only few
optical lines) and their winds, also the UV ($\rightarrow$ P~Cygni
profiles) might be required. Moreover, an analysis of the outer wind
requires the knowledge of the mm and radio regime (free-free
emission), and constraints for the shock emission are found from the
X-ray band, which needs to be known as well for the analysis of
stellar coronae and compact objects.

In addition to spectroscopy, all other astronomical techniques such as
interferometry (stellar and wind shape, binarity), polarimetry
(inhomogeneities, wind and disk shape, magnetic fields), imaging
(stellar surroundings), photometry (integrated fluxes), and astrometry
(distances, peculiar velocities, and proper motions) are needed
to obtain a consistent picture. Particularly, the combination of
photometry and astrometry is inevitable, since otherwise there would
be no handle on the stellar radius, which results from a comparison of
apparent ($\rightarrow$ photometry) and absolute ($\rightarrow$ models
+ distance) fluxes, corrected for reddening.

Time series are required to study the presence of companions (stars,
exoplanets), temporary structures (spots, narrow absorption
components), and the evolution of transient phenomena (outburst,
mergers, supernovae). The analysis of pulsation patterns present in
all stars from spectroscopic and/or photometric time series has
opened the way to the unseen subphotospheric layers and the deep
stellar interiors.

\subsection{Numerical Codes}
\label{codes} 
Because of the different physical conditions as a function of stellar
spectral type and evolutionary phase, we split the various codes into
the scheme LTE vs. NLTE and 1-D vs. 3-D. In
Table~\ref{code_comparison}, we enumerate only those codes that
are/have been frequently used (without aiming at completeness). LTE
codes usually include molecules, and 1-D structure codes base on the
mixing length theory if convection plays a role. For a detailed
description and references, we refer to
\citet[Chap.~3]{Crivellari2019}, except for ASSET (see
\citealt{Koesterke2009}) and for SYNPLE (see {\tt
github.com/callendeprieto/synple}). Particularly for the LTE case,
many codes provide only an atmospheric model (i.e., density and
temperature structure), whilst the corresponding SED (in absolute
flux-units and/or normalized to the neighbouring continuum) is
synthesized by a separate, ``diagnostic'' code, partly by
(re-)calculating the occupation numbers (sometimes in NLTE, which
constitutes a hybrid approach).  If, on the other hand, the occupation
numbers from the atmospheric model were used, this second step is
called the \textit{formal integral}, since then ``only'' the (mostly
steady-state) equation of RT (Eq.~\ref{eq_RT}) needs to be solved. In
most NLTE approaches, atmosphere and formal integral are still
calculated in two steps, but are provided in one common package.  

\begin{table}[t]
\caption{Various numerical codes with different assumptions and
purposes: AS -- atmospheric structure code; DIA -- diagnostic code (see
text); FI -- formal integral; RMHD -- radiation magneto-hydrodynamics
code. When not stated differently, and except for RMHD codes, 
stationary approach used. Abbreviations for features: hs -- hydrostatic;
hd -- hydrodynamic; unif -- unified (photosphere+wind, smoothly
connected); pp -- plane-parallel; ss -- spherically symmetric; RE --
radiative equilibrium; MLT -- mixing length theory; EB -- electron thermal
balance (see \citealt{Hummer1963, Kubat1999}); cart -- Cartesian; sph -- spherical; cyl -- cylindrical}
\label{code_comparison}
\tabcolsep1.5mm
\begin{center}
\small{ 
\begin{tabular}{llllll}
\hline
   &  purpose &  name  & main author(s)  & features  & comments \\
\hline
\multirow{6}{*}{LTE 1-D} & \multirow{3}{*}{AS} & ATLAS9 & 
\multirow{2}{*}{R. Kurucz}  & \multirow{2}{*}{hs, pp, RE+MLT} & opacity distribution functions \\ 
        &                & ATLAS12&    &        & opacity sampling\\
        &                & MARCS  & B. Gustafsson, B. Plez & hs, pp/ss, RE+MLT&
opacities interpolated from pre-computed tables \\
\cline{2-6}
        & \multirow{3}{*}{DIA} & SYNSPEC & I. Hubený, T. Lanz & & 
	\multirow{2}{*}{used in combination with ATLAS, TLUSTY(LTE)}\\
        &                & SYNTHE  & R. Kurucz  & & \\
        &                & Turbospectrum & B. Plez & & used in
	combination with MARCS, ATLAS, {\ldots} \\
\hline
LTE 3-D & DIA & ASSET  & L. Koesterke &   &  SEDs from 3-D (RM)HD models \\
\hline
\multirow{9}{*}{NLTE 1-D} & \multirow{5}{*}{AS+FI} & CMFGEN &
 J. Hillier, L. Dessart & unif, ss, RE &complete CMF transfer$^a$ \\
&  & FASTWIND  & J. Puls, E. Santolaya-Rey & unif, ss, EB &CMF transfer
for individual elements$^b$ \\
&  &  PHOENIX & P. Hauschildt, E. Baron & unif, ss, RE+MLT & complete CMF transfer$^c$\\ 
&  &  PoWR & W.-R. Hamann & unif, ss, RE & complete CMF transfer$^d$\\ 
&  &  WM-basic & A. Pauldrach, T. Hoffmann & hd (stat.), ss, EB & Sobolev line
transfer$^e$\\ 
\cline{2-6}
& AS & TLUSTY & I. Hubený, T. Lanz & hs, pp, RE & used in combination
with SYNSPEC \\
\cline{2-6}
& \multirow{4}{*}{DIA} & SYNSPEC & I. Hubený, T. Lanz & & used in combination
with TLUSTY\\
& & SYNPLE & C. Allende Prieto&  & python wrapper for SYNSPEC (LTE \&
NLTE) \\
& & \sc{detail/surface} & K. Butler, J. Giddings & & used in combination with
ATLAS$^f$ or TLUSTY\\
& & Multi & M. Carlsson & & used in combination with external
atm. codes$^g$\\
\hline
\multirow{3}{*}{NLTE 3-D} & AS+FI & PHOENIX/3D  & P. Hauschildt, E. Baron & unified, RE+MLT 
& cart/sph/cyl coordinate systems possible\\
\cline{2-6}
& \multirow{2}{*}{DIA} & Multi3d & J. Leenaarts et al.$^h$ & &
used in combination with external 3-D atm. codes$^i$\\
& & RH & H. Uitenbroek & & used in combination with external 3-D atm. codes$^j$\\ 
\hline
\multirow{3}{*}{RMHD} & \multirow{3}{*}{AS} & CO5BOLD & B. Freytag, M.
Steffen & hd (time-dep.), cart & see$^m$\\
& & Bifrost &  B. Gudiksen et al.$^k$ & hd (time-dep.), cart & see$^m$ \\
& &  Stagger & \AA. Nordlund et al.$^l$ & hd (time-dep.), cart & see$^m$ \\ 
\hline
\end{tabular}
}
\begin{tablenotes}
\footnotesize{
In the following, ``optically thick winds'' 
refer to high mass-loss rates where $\tau_{\rm R} = 1$ is
reached within the wind, e.g., winds from classical Wolf-Rayet stars\\
$^a$ optically thin and thick winds. Time-dependent calculations for 
SN-remnants also possible\\
$^b$ optically thin winds only. Fast computation time because of specific
treatment of background elements. Optically thick clumping included in NLTE 
calculations and formal integral. Version 11 (in work) allows for complete CMF
transfer\\
$^c$ optically thin and thick winds. Molecules included, SN-remnants
(stationary) possible --  
$^d$ optically thin and thick winds. Optically thick clumping in 
formal integral\\
$^e$ optically thin winds only. SN-remnants (stationary) possible. No
clumping included -- 
$^f$ hybrid approach, see main text\\
$^g$ one single atom in NLTE, others in LTE or read in --
$^h$ CoIs: J. Bj{\o}rgen, A. Sukhorukov \\
$^i$ one single atom in NLTE, others in LTE; hydrogen population can
be read in --
$^j$ multiple atoms and molecules\\
$^k$ CoIs: M. Carlsson, V. Hansteen, J. Leenaarts, J. Martinez-Sykora --
$^l$ CoIs: K. Galsgaard, R. Collet, R. Stein \\
$^m$ used to model various RMHD problems related to 
stellar evenvelopes (and other realms) such as convection and turbulence. 
The output of these models (either directly or suitable averages) can
be used to synthesize corresponding SEDs.\\
}
\end{tablenotes}
\end{center}
\end{table}

\subsection{Determination of Stellar and Wind Parameters}
\label{methods}
\begin{figure}[t]
\centering \includegraphics[width=0.7\textwidth]{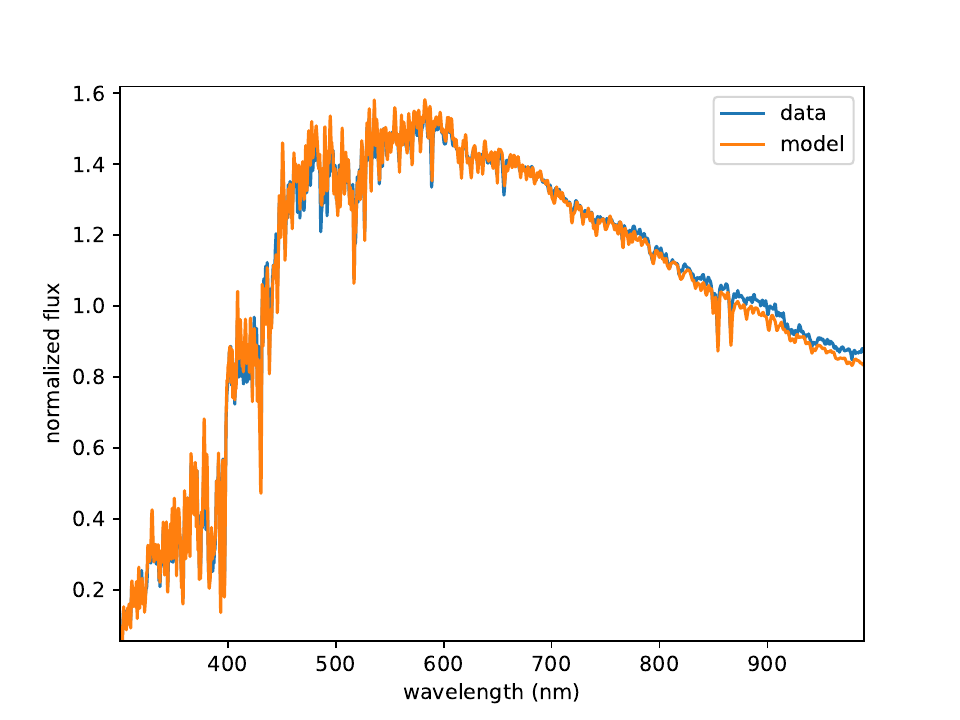}
\caption {Observed \citep{Bohlin2019} and synthetic best-fit optical
spectrum for the K 0.5 III star 2MASS J17551622+6610116. 
Synthetic spectra computed with SYNSPEC based on a Kurucz model.
Derived parameters: \Teff\ $\approx$ 4870~K, \logg\ = 3.0 (cgs), and solar 
iron abundance.
Fitting method: Bayesian algorithm in Synple (see Sect.~\ref{codes}).
}
\label{spec_cool}
\end{figure}

\begin{figure}[t]
\centering \includegraphics[width=0.7\textwidth]{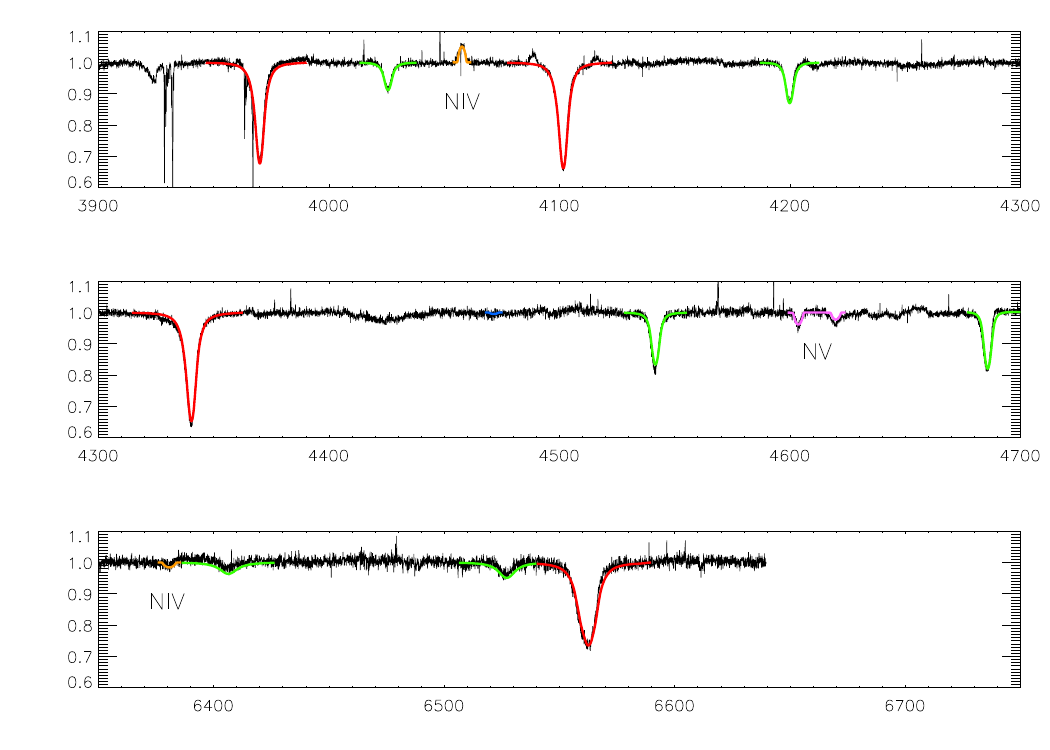}
\caption {Observed and synthetic best-fit optical spectrum for the O2
dwarf BI237 in the Large Magellanic Cloud. Observations from
\citet{Doran2011}, synthetic spectra for H, He, and N from FASTWIND:
\Teff\ $\approx$ 53,000~K, \logg\ = 4.1, low \mdot, nitrogen mildly
enriched (see \citealt{Rivero2012}). Fitting method: by-eye. Profiles
in red: \HI; in blue: \HeI\ (one very weak line); in green: \HeII; in
orange: \NIV; and in magenta \NV. Note the few number of metal lines
being present in the optical range. The emission in \NIV~4058\,$\AA$
is a combined wind and NLTE effect.} 
\label{spec_O2}
\end{figure}

The basic idea of quantitative spectroscopy is to compare (by various
methods, outlined below) observed and theoretical spectra/SEDs, where
the input parameters for the atmospheric models are modified until
synthetic and observed spectra/SEDs agree. In a first step, the
stellar (and wind parameters) need to be quantified, before individual
abundances are derived in a second step. To this end, and in
dependence of spectral type, different observed diagnostic features
are compared with model spectra\footnote{after accounting for
appropriate rotational ($v_{\rm rot} \sin i$), macroturbulent ($v_{\rm
mac}$) and instrumental broadening, via corresponding convolutions} to
infer \Teff, \logg\ (or $\log g_{\rm eff}$, see Sect.~\ref{rot}), and
often the helium content in hot stars (being the second most abundant
element, changes in its abundance may modify the atmospheric
structure). The effective temperature might, for cooler stars
(spectral types F and later), be derived from photometry, or from
specific diagnostic features such as the collisionally-broadened wings
of hydrogen lines, or by requiring atomic iron lines with various
excitation energies to give the same iron abundance. In early-type
stars, the ionization equilibrium (lines from different ions of the
same element shall be reproduced with a similar quality) is exploited
(nitrogen, helium, silicon and magnesium ionization equilibrium for
hot and cooler O-stars, B-stars, and A-stars, respectively). The
(effective) gravity in cooler stars is derived from the damping wings
of metal lines, broadened mainly by collisions with hydrogen atoms, or
by imposing the ionization balance for atomic and ionized iron. In
hotter stars, it is determined from the Stark-broadened wings of \Hg\
and \Hd, which are predominantely formed in the photosphere and
strongly react on the electron-density controlled by $\log g_{\rm
(eff)}$ (see Eq.~\ref{eq_scaleheight}).  In massive stars with a
considerable wind strength, many photospheric lines become
contaminated by wind-effects (see also Sect.~\ref{winds}), and thus
the wind-parameters (mass-loss rate, terminal velocity and
parameterized velocity-field, plus specific parameters describing the
distribution of inhomogeneities) need to be inferred in parallel,
mostly from \Ha, \HeII\,4686\,\AA, and, if available, the ultraviolet
P Cygni lines. Further details can be found in \citet{Gray2021} for
stars of later spectral type, and in \citet{SSD2020} for early type
stars. Examples for the parameter determination for a cool and a hot
star from optical spectra alone are presented in Fig.~\ref{spec_cool}
and ~\ref{spec_O2}, respectively.
%fig_spec_cool
%
%fig_spec_O2

\subsection{The Stellar Mass}\label{masses}
The stellar mass is the primary parameter determining the structure
and fate of a star. The best method to obtain stellar masses is to
observe and analyze binary stars, and eclipsing binaries in
particular. Subject to their mutual gravitational influence, their
motion is determined by their respective masses (and orbital
separation). For low mass stars, the observation of binaries allows a
precise comparison with the theory of stellar structure and 
evolution, as there is a large number of systems with components that
do not interact and behave as isolated stars. The situation is much
more complicated for high-mass stars: gravitational fields are more
intense, separations smaller, interaction effects stronger and the
number of available systems comparatively small. This stresses
the importance to obtain masses from high-quality spectroscopic
analyses: After correcting the measured effective gravity for the
centrifugal acceleration (Sect.~\ref{rot}), a combination of \logg\
with the stellar radius (Sect.~\ref{obs}) yields the so-called
\textit{spectroscopic mass}, which is subject to significant
uncertainties, mainly linked to the determination of radius and
gravity. Though these uncertainties accumulate into relatively large
errors for the individual masses (sometimes exceeding 50\%), a
so-called mass discrepancy has been identified \citep{Herrero1992}, a
systematic difference between spectroscopic and \textit{evolutionary
masses} \footnote{inferred from evolutionary calculations, as a
function of luminosity (from \Teff\ and \Rstar) and effective
temperature (theoretical Hertzsprung-Russel diagram)} of single stars.

In spite of some decades of efforts, this discrepancy has not been
fully resolved, with varying results when different samples have been
analyzed. The comparison between both mass determinations has become
more complicated after the realization that many massive stars are in
binary systems or have suffered different degrees of interaction in
their evolution. Beyond the Magellanic Clouds, crowding enhances the
difficulties as undetected visual companions may contaminate the
spectrum and increase the uncertainties. Clearly, reducing the
uncertainty in the determination of stellar masses of massive stars
from quantitative spectroscopy via model atmospheres will be a major
leap forward.

\subsection{Determination of Abundances}
\label{abund}
When the stellar (and wind) parameters have been determined, 
individual abundances for specific elements can be derived, as long as
corresponding spectral lines are visible (although a lack of spectral
lines can also be used to constrain the abundance of a given element).
Two different methods might be employed: (i) similar to the procedure
outlined above, the observed line profiles are fitted by synthetic
ones, where now, for fixed stellar (and wind) parameters, the input
values for the abundances are varied, in parallel with the
micro-turbulent velocity (Sect.~\ref{inhomo}). A certain disadvantage
of this method is its dependence on quite precise values for $v_{\rm
rot} \sin i$ and $v_{\rm mac}$. (ii) Alternatively, not the profile
shape, but the equivalent width (see Sect.~\ref{rot}) is fitted, which
avoids the previous problem since the equivalent width is conserved by
both broadening processes. This, in turn, has the disadvantage of a
certain degeneracy; for example, the line core could be underestimated
and the wings overestimated, still matching the equivalent width.
Also here, $v_{\rm mic}$ needs to be determined in parallel.
Generally, a reasonable solution requires to find the same abundance
within any diagnostic line from the atom under consideration (in
particular, for lines from different ions, if present), together with
a unique value for $v_{\rm mic}$. Obviously, the more lines are
available, the easier it is to discriminate between systematic and
statistical errors. Moreover, it is useful to concentrate on weak and
intermediate strong lines (where the profile-strength and equivalent
width react linearly on the abundance), to avoid saturation effects
and to minimize the role of microturbulence.

\subsection{Fitting Methods}
\label{fitting}
It remains to be outlined how the above fitting process is actually
performed. In most cases, the user needs to define a merit-function,
which typically consists of a (weighted) $\chi^2$, measuring the sum
of squared differences between synthetic and observed fluxes/profiles
as a function of wavelength, normalized to the variance due to photon
noise (accounting for systematics as well). The best fitting
parameters then result from that model which produces the minimum
$\chi^2$, and the errors for the individual parameters can be found
from the $\chi^2$ distribution around that minimum. There are various
methods to find the global minimum, and all of them are still applied.
The oldest method is a by-eye inspection of the fit-quality using a
multitude of models, which for a large parameter space (when also
wind-parameters need to be obtained) becomes cumbersome (besides being
more subjective). A better method is to perform the fit using
pre-calculated model grids (partly of quite large dimension, e.g.
\citealt{2006ApJ...636..804A, 2016AJ....151..144G, Holgado2018}), or
to use genetic algorithms if it is possible to calculate new models on
the fly (e.g., \citealt{Mokiem2005}). In recent years, MCMC approaches
(which directly deliver the probability density distribution for the
analyzed parameters) and machine-learning methods have become popular,
as well as data-driven models \citep{2015ApJ...808...16N,
2019ApJ...879...69T}. 

\subsection{Error Budget}
\label{errors}
Typical errors for \Teff\ and \logg\ are 3-5\% and 0.1 dex (but see
above), respectively. Mass-loss rates for thinner winds (where \Ha\ is
still in absorption) can be determined only with a precision of a
factor of two, because of an uncertain velocity field. For objects
with larger \mdot, the precision can be higher (up to roughly 10\%), 
but only if the clumping-stratification can be
accurately derived, which is difficult and requires a multi-wavelength
analysis. Typical errors for the Helium abundance are 10 to 30\%,
and for the stellar radius 5 to 15\% (if reliable distances and
reddening properties are available). Regarding individual abundances
except for He, there is usually a large difference between the quoted
errors for O-type stars and cooler ones. Since in most cases the
abundances have been derived for fixed stellar parameters, a careful
error propagation would be necessary, though often not done. Moreover,
the number of visible metallic lines in O-stars is much less than in
cooler ones, and the occupation numbers are often strongly affected by
subtle NLTE-effects (see, e.g., \citealt{Puls2020} for the case of
nitrogen), which might lead to significant systematics. Overall, in
stars later than O the abundance errors are on the order of 10\%,
whilst in O-stars they might reach 0.1 up to 0.2 or even 0.3 dex,
except for Helium. We stress that for those elements with extremely
few visible lines (e.g., lithium and boron), the error budget is
strongly dependent on the precision of the atomic data. Larger errors
also result when lines from only one ionization stage were present,
since then a consistency check on a correct ionization equilibrium is
not possible. 

\section{Summary and Conclusions}\label{summary} 
In this chapter, we have summarized the basic theories and approaches
to model stellar atmospheres and corresponding synthetic spectra,
always discriminating between the pecularities of earlier and laters
spectral types. We discussed as well the physics and influence of
(various kinds of) stellar winds, rotation, magnetic fields,
inhomogeneities, and multiplicity. We stressed the impact of multi-D
RMHD models to further our understanding of cool-star atmospheres
(e.g., convection) and hot star winds (and atmospheres),
reproducing many features in the observed spectra (and images, when
considering the Sun). Stellar envelopes are an
ideal laboratory to investigate specific physical effects under
extreme conditions, such as the line-deshadowing instability and
corresponding radiative-accoustic waves. We finalized this chapter by
describing the methodology to analyze observed SEDs by means of
quantitative spectroscopy, and referred to the so-called
mass-discrepancy found in the massive star domain. As a last comment,
we like to warn any potential user about a black-box usage of model
atmosphere codes (at least as long as specific physical problems have
not been settled), and to be aware of the underlying and/or adopted
physics.

\begin{ack}[Acknowledgments] 

JP gratefully acknowledges the stimulating atmosphere (sic!) at the
LMU University Observatory during his recent 40 years of scientific
work and lecturing, as well as the fruitful collaboration with
colleagues and students from all over the world, in particular R.-P.
Kudritzki, S. Owocki, and the late A.W.A. Pauldrach and D.G. Hummer.
AH acknowledges the many collaborations with colleagues at the LMU on
stellar atmospheres, as well as support from the Spanish Ministry of
Science and Innovation (MICINN) through the Spanish State Research
Agency through grants PID2021-122397NB-C21, and the Severo Ochoa
Programme 2020-2023 (CEX2019-000920-S). CAP is thankful to his
lifelong collaborators Y. Osorio, H. Ludwig, L. Koesterke, P. Barklem,
M. Asplund, D. Lambert, B. Ruiz-Cobo, and I. Hubeny in particular, for
their help, support and so many exciting discussions.

\end{ack}

\seealso{\cite{Payne1925}, \cite{Unsold1955},
\cite{Aller1963}, \cite{Mihalas1978}, \cite{Gray2021}, 
\cite[H\&M]{Hubeny2015}, \cite{Crivellari2019}, \cite{SSD2020}}

\bibliographystyle{Harvard}
\bibliography{reference_stellar_atm}

\begin{thebibliography*}{93}
\providecommand{\bibtype}[1]{}
\providecommand{\natexlab}[1]{#1}
{\catcode`\|=0\catcode`\#=12\catcode`\@=11\catcode`\\=12
|immediate|write|@auxout{\expandafter\ifx\csname
  natexlab\endcsname\relax\gdef\natexlab#1{#1}\fi}}
\renewcommand{\url}[1]{{\tt #1}}
\providecommand{\urlprefix}{URL }
\expandafter\ifx\csname urlstyle\endcsname\relax
  \providecommand{\doi}[1]{doi:\discretionary{}{}{}#1}\else
  \providecommand{\doi}{doi:\discretionary{}{}{}\begingroup
  \urlstyle{rm}\Url}\fi
\providecommand{\bibinfo}[2]{#2}
\providecommand{\eprint}[2][]{\url{#2}}

\bibtype{Article}%
\bibitem[{Abdul-Masih} et al.(2020)]{Abdul-Masih2020}
\bibinfo{author}{{Abdul-Masih} M}, \bibinfo{author}{{Sana} H},
  \bibinfo{author}{{Conroy} KE}, \bibinfo{author}{{Sundqvist} J},
  \bibinfo{author}{{Pr{\v{s}}a} A}, \bibinfo{author}{{Kochoska} A} and
  \bibinfo{author}{{Puls} J} (\bibinfo{year}{2020}), \bibinfo{month}{Apr.}
\bibinfo{title}{{Spectroscopic patch model for massive stars using PHOEBE II
  and FASTWIND}}.
\bibinfo{journal}{{\em \aap}} \bibinfo{volume}{636}, \bibinfo{eid}{A59}.
  \bibinfo{doi}{\doi{10.1051/0004-6361/201937341}}.
\eprint{2003.09008}.

\bibtype{Article}%
\bibitem[{Aerts} et al.(2009)]{Aerts2009}
\bibinfo{author}{{Aerts} C}, \bibinfo{author}{{Puls} J},
  \bibinfo{author}{{Godart} M} and  \bibinfo{author}{{Dupret} MA}
  (\bibinfo{year}{2009}), \bibinfo{month}{Dec.}
\bibinfo{title}{{Collective pulsational velocity broadening due to gravity
  modes as a physical explanation for macroturbulence in hot massive stars}}.
\bibinfo{journal}{{\em \aap}} \bibinfo{volume}{508} (\bibinfo{number}{1}):
  \bibinfo{pages}{409--419}. \bibinfo{doi}{\doi{10.1051/0004-6361/200810471}}.
\eprint{0909.3585}.

\bibtype{Article}%
\bibitem[{Allende Prieto}(2023)]{2023Atoms..11...61A}
\bibinfo{author}{{Allende Prieto} C} (\bibinfo{year}{2023}),
  \bibinfo{month}{Mar.}
\bibinfo{title}{{The Shapes of Stellar Spectra}}.
\bibinfo{journal}{{\em Atoms}} \bibinfo{volume}{11} (\bibinfo{number}{3}),
  \bibinfo{eid}{61}. \bibinfo{doi}{\doi{10.3390/atoms11030061}}.
\eprint{2303.14340}.

\bibtype{Article}%
\bibitem[{Allende Prieto} et al.(2006)]{2006ApJ...636..804A}
\bibinfo{author}{{Allende Prieto} C}, \bibinfo{author}{{Beers} TC},
  \bibinfo{author}{{Wilhelm} R}, \bibinfo{author}{{Newberg} HJ},
  \bibinfo{author}{{Rockosi} CM}, \bibinfo{author}{{Yanny} B} and
  \bibinfo{author}{{Lee} YS} (\bibinfo{year}{2006}), \bibinfo{month}{Jan.}
\bibinfo{title}{{A Spectroscopic Study of the Ancient Milky Way: F- and G-Type
  Stars in the Third Data Release of the Sloan Digital Sky Survey}}.
\bibinfo{journal}{{\em \apj}} \bibinfo{volume}{636} (\bibinfo{number}{2}):
  \bibinfo{pages}{804--820}. \bibinfo{doi}{\doi{10.1086/498131}}.
\eprint{astro-ph/0509812}.

\bibtype{Book}%
\bibitem[{Aller}(1963)]{Aller1963}
\bibinfo{author}{{Aller} LH} (\bibinfo{year}{1963}).
\bibinfo{title}{{Astrophysics. The atmospheres of the sun and stars}}.

\bibtype{Article}%
\bibitem[{Asplund} et al.(2009)]{Asplund2009}
\bibinfo{author}{{Asplund} M}, \bibinfo{author}{{Grevesse} N},
  \bibinfo{author}{{Sauval} AJ} and  \bibinfo{author}{{Scott} P}
  (\bibinfo{year}{2009}), \bibinfo{month}{Sep.}
\bibinfo{title}{{The Chemical Composition of the Sun}}.
\bibinfo{journal}{{\em \araa}} \bibinfo{volume}{47} (\bibinfo{number}{1}):
  \bibinfo{pages}{481--522}.
  \bibinfo{doi}{\doi{10.1146/annurev.astro.46.060407.145222}}.
\eprint{0909.0948}.

\bibtype{Article}%
\bibitem[{Auer} and {Mihalas}(1969)]{Auer1969}
\bibinfo{author}{{Auer} LH} and  \bibinfo{author}{{Mihalas} D}
  (\bibinfo{year}{1969}), \bibinfo{month}{Nov.}
\bibinfo{title}{{Non-Lte Model Atmospheres. III. a Complete-Linearization
  Method}}.
\bibinfo{journal}{{\em \apj}} \bibinfo{volume}{158}: \bibinfo{pages}{641}.
  \bibinfo{doi}{\doi{10.1086/150226}}.

\bibtype{Article}%
\bibitem[{Barklem}(2016)]{2016PhRvA..93d2705B}
\bibinfo{author}{{Barklem} PS} (\bibinfo{year}{2016}), \bibinfo{month}{Apr.}
\bibinfo{title}{{Excitation and charge transfer in low-energy hydrogen-atom
  collisions with neutral atoms: Theory, comparisons, and application to Ca}}.
\bibinfo{journal}{{\em \pra}} \bibinfo{volume}{93} (\bibinfo{number}{4}),
  \bibinfo{eid}{042705}. \bibinfo{doi}{\doi{10.1103/PhysRevA.93.042705}}.
\eprint{1603.07097}.

\bibtype{Article}%
\bibitem[{Barklem}(2018)]{2018A&A...610A..57B}
\bibinfo{author}{{Barklem} PS} (\bibinfo{year}{2018}), \bibinfo{month}{Feb.}
\bibinfo{title}{{Excitation and charge transfer in low-energy hydrogen atom
  collisions with neutral oxygen}}.
\bibinfo{journal}{{\em \aap}} \bibinfo{volume}{610}, \bibinfo{eid}{A57}.
  \bibinfo{doi}{\doi{10.1051/0004-6361/201731968}}.
\eprint{1712.01166}.

\bibtype{Article}%
\bibitem[{Barklem} et al.(2012)]{2012A&A...541A..80B}
\bibinfo{author}{{Barklem} PS}, \bibinfo{author}{{Belyaev} AK},
  \bibinfo{author}{{Spielfiedel} A}, \bibinfo{author}{{Guitou} M} and
  \bibinfo{author}{{Feautrier} N} (\bibinfo{year}{2012}), \bibinfo{month}{May}.
\bibinfo{title}{{Inelastic Mg+H collision data for non-LTE applications in
  stellar atmospheres}}.
\bibinfo{journal}{{\em \aap}} \bibinfo{volume}{541}, \bibinfo{eid}{A80}.
  \bibinfo{doi}{\doi{10.1051/0004-6361/201219081}}.
\eprint{1203.4877}.

\bibtype{Article}%
\bibitem[{Barklem} et al.(2017)]{2017A&A...606A..11B}
\bibinfo{author}{{Barklem} PS}, \bibinfo{author}{{Osorio} Y},
  \bibinfo{author}{{Fursa} DV}, \bibinfo{author}{{Bray} I},
  \bibinfo{author}{{Zatsarinny} O}, \bibinfo{author}{{Bartschat} K} and
  \bibinfo{author}{{Jerkstrand} A} (\bibinfo{year}{2017}),
  \bibinfo{month}{Sep.}
\bibinfo{title}{{Inelastic e+Mg collision data and its impact on modelling
  stellar and supernova spectra}}.
\bibinfo{journal}{{\em \aap}} \bibinfo{volume}{606}, \bibinfo{eid}{A11}.
  \bibinfo{doi}{\doi{10.1051/0004-6361/201730864}}.
\eprint{1706.03399}.

\bibtype{Article}%
\bibitem[{Bergemann} et al.(2012)]{Bergemann2012}
\bibinfo{author}{{Bergemann} M}, \bibinfo{author}{{Lind} K},
  \bibinfo{author}{{Collet} R}, \bibinfo{author}{{Magic} Z} and
  \bibinfo{author}{{Asplund} M} (\bibinfo{year}{2012}), \bibinfo{month}{Nov.}
\bibinfo{title}{{Non-LTE line formation of Fe in late-type stars - I. Standard
  stars with 1D and <3D> model atmospheres}}.
\bibinfo{journal}{{\em \mnras}} \bibinfo{volume}{427} (\bibinfo{number}{1}):
  \bibinfo{pages}{27--49}.
  \bibinfo{doi}{\doi{10.1111/j.1365-2966.2012.21687.x}}.
\eprint{1207.2455}.

\bibtype{Article}%
\bibitem[{Bohlin} et al.(2019)]{Bohlin2019}
\bibinfo{author}{{Bohlin} RC}, \bibinfo{author}{{Deustua} SE} and
  \bibinfo{author}{{de Rosa} G} (\bibinfo{year}{2019}), \bibinfo{month}{Nov.}
\bibinfo{title}{{Hubble Space Telescope Flux Calibration. I. STIS and
  CALSPEC}}.
\bibinfo{journal}{{\em \aj}} \bibinfo{volume}{158} (\bibinfo{number}{5}),
  \bibinfo{eid}{211}. \bibinfo{doi}{\doi{10.3847/1538-3881/ab480c}}.

\bibtype{Article}%
\bibitem[{Bondi}(1952)]{Bondi1952}
\bibinfo{author}{{Bondi} H} (\bibinfo{year}{1952}), \bibinfo{month}{Jan.}
\bibinfo{title}{{On spherically symmetrical accretion}}.
\bibinfo{journal}{{\em \mnras}} \bibinfo{volume}{112}: \bibinfo{pages}{195}.
  \bibinfo{doi}{\doi{10.1093/mnras/112.2.195}}.

\bibtype{Article}%
\bibitem[{Brands} et al.(2022)]{Brands2022}
\bibinfo{author}{{Brands} SA}, \bibinfo{author}{{de Koter} A},
  \bibinfo{author}{{Bestenlehner} JM}, \bibinfo{author}{{Crowther} PA},
  \bibinfo{author}{{Sundqvist} JO}, \bibinfo{author}{{Puls} J},
  \bibinfo{author}{{Caballero-Nieves} SM}, \bibinfo{author}{{Abdul-Masih} M},
  \bibinfo{author}{{Driessen} FA}, \bibinfo{author}{{Garc{\'\i}a} M},
  \bibinfo{author}{{Geen} S}, \bibinfo{author}{{Gr{\"a}fener} G},
  \bibinfo{author}{{Hawcroft} C}, \bibinfo{author}{{Kaper} L},
  \bibinfo{author}{{Keszthelyi} Z}, \bibinfo{author}{{Langer} N},
  \bibinfo{author}{{Sana} H}, \bibinfo{author}{{Schneider} FRN},
  \bibinfo{author}{{Shenar} T} and  \bibinfo{author}{{Vink} JS}
  (\bibinfo{year}{2022}), \bibinfo{month}{Jul.}
\bibinfo{title}{{The R136 star cluster dissected with Hubble Space
  Telescope/STIS. III. The most massive stars and their clumped winds}}.
\bibinfo{journal}{{\em \aap}} \bibinfo{volume}{663}, \bibinfo{eid}{A36}.
  \bibinfo{doi}{\doi{10.1051/0004-6361/202142742}}.
\eprint{2202.11080}.

\bibtype{Article}%
\bibitem[{Caffau} et al.(2011)]{2011SoPh..268..255C}
\bibinfo{author}{{Caffau} E}, \bibinfo{author}{{Ludwig} HG},
  \bibinfo{author}{{Steffen} M}, \bibinfo{author}{{Freytag} B} and
  \bibinfo{author}{{Bonifacio} P} (\bibinfo{year}{2011}), \bibinfo{month}{Feb.}
\bibinfo{title}{{Solar Chemical Abundances Determined with a CO5BOLD 3D Model
  Atmosphere}}.
\bibinfo{journal}{{\em \solphys}} \bibinfo{volume}{268} (\bibinfo{number}{2}):
  \bibinfo{pages}{255--269}. \bibinfo{doi}{\doi{10.1007/s11207-010-9541-4}}.
\eprint{1003.1190}.

\bibtype{Inproceedings}%
\bibitem[{Cantiello} et al.(2011)]{Cantiello2011}
\bibinfo{author}{{Cantiello} M}, \bibinfo{author}{{Braithwaite} J},
  \bibinfo{author}{{Brandenburg} A}, \bibinfo{author}{{Del Sordo} F},
  \bibinfo{author}{{K{\"a}pyl{\"a}} P} and  \bibinfo{author}{{Langer} N}
  (\bibinfo{year}{2011}), \bibinfo{month}{Jul.}, \bibinfo{title}{{3D MHD
  simulations of subsurface convection in OB stars}}, \bibinfo{editor}{{Neiner}
  C}, \bibinfo{editor}{{Wade} G}, \bibinfo{editor}{{Meynet} G} and
  \bibinfo{editor}{{Peters} G}, (Eds.), \bibinfo{booktitle}{Active OB Stars:
  Structure, Evolution, Mass Loss, and Critical Limits}, \bibinfo{volume}{272},
   \bibinfo{pages}{32--37}, \eprint{1009.4462}.

\bibtype{Article}%
\bibitem[{Castor} et al.(1975)]{CAK1975}
\bibinfo{author}{{Castor} JI}, \bibinfo{author}{{Abbott} DC} and
  \bibinfo{author}{{Klein} RI} (\bibinfo{year}{1975}), \bibinfo{month}{Jan.}
\bibinfo{title}{{Radiation-driven winds in Of stars.}}
\bibinfo{journal}{{\em \apj}} \bibinfo{volume}{195}: \bibinfo{pages}{157--174}.
  \bibinfo{doi}{\doi{10.1086/153315}}.

\bibtype{Article}%
\bibitem[{Cowley}(1990)]{Cowley1990}
\bibinfo{author}{{Cowley} CR} (\bibinfo{year}{1990}), \bibinfo{month}{Jan.}
\bibinfo{title}{{Second Viscosity of the Gas in the Outer Solar Envelope}}.
\bibinfo{journal}{{\em \apj}} \bibinfo{volume}{348}: \bibinfo{pages}{328}.
  \bibinfo{doi}{\doi{10.1086/168239}}.

\bibtype{Article}%
\bibitem[{Cranmer} and {Owocki}(1995)]{Cranmer1995}
\bibinfo{author}{{Cranmer} SR} and  \bibinfo{author}{{Owocki} SP}
  (\bibinfo{year}{1995}), \bibinfo{month}{Feb.}
\bibinfo{title}{{The Effect of Oblateness and Gravity Darkening on the
  Radiation Driving in Winds from Rapidly Rotating B Stars}}.
\bibinfo{journal}{{\em \apj}} \bibinfo{volume}{440}: \bibinfo{pages}{308}.
  \bibinfo{doi}{\doi{10.1086/175272}}.

\bibtype{Book}%
\bibitem[{Crivellari} et al.(2019)]{Crivellari2019}
\bibinfo{author}{{Crivellari} L}, \bibinfo{author}{{Sim{\'o}n-D{\'\i}az} S} and
   \bibinfo{author}{{Ar{\'e}valo} MJ} (\bibinfo{year}{2019}).
\bibinfo{title}{{Radiative Transfer in Stellar and Planetary Atmospheres}}.
\bibinfo{doi}{\doi{10.1017/9781108583572}}.

\bibtype{Article}%
\bibitem[{Debnath} et al.(2024)]{Debnath2024}
\bibinfo{author}{{Debnath} D}, \bibinfo{author}{{Sundqvist} JO},
  \bibinfo{author}{{Moens} N}, \bibinfo{author}{{Van der Sijpt} C},
  \bibinfo{author}{{Verhamme} O} and  \bibinfo{author}{{Poniatowski} LG}
  (\bibinfo{year}{2024}), \bibinfo{month}{Apr.}
\bibinfo{title}{{2D unified atmosphere and wind simulations of O-type stars}}.
\bibinfo{journal}{{\em \aap}} \bibinfo{volume}{684}, \bibinfo{eid}{A177}.
  \bibinfo{doi}{\doi{10.1051/0004-6361/202348206}}.
\eprint{2401.08391}.

\bibtype{Article}%
\bibitem[{Doran} and {Crowther}(2011)]{Doran2011}
\bibinfo{author}{{Doran} EI} and  \bibinfo{author}{{Crowther} PA}
  (\bibinfo{year}{2011}), \bibinfo{month}{Jan.}
\bibinfo{title}{{A VLT/UVES spectroscopy study of O2 stars in the LMC}}.
\bibinfo{journal}{{\em Bulletin de la Societe Royale des Sciences de Liege}}
  \bibinfo{volume}{80}: \bibinfo{pages}{129--133}.

\bibtype{Article}%
\bibitem[{Feldmeier}(1995)]{Feldmeier1995}
\bibinfo{author}{{Feldmeier} A} (\bibinfo{year}{1995}), \bibinfo{month}{Jul.}
\bibinfo{title}{{Time-dependent structure and energy transfer in hot star
  winds.}}
\bibinfo{journal}{{\em \aap}} \bibinfo{volume}{299}: \bibinfo{pages}{523}.

\bibtype{Article}%
\bibitem[{Freytag} and {H{\"o}fner}(2023)]{Freytag2023}
\bibinfo{author}{{Freytag} B} and  \bibinfo{author}{{H{\"o}fner} S}
  (\bibinfo{year}{2023}), \bibinfo{month}{Jan.}
\bibinfo{title}{{Global 3D radiation-hydrodynamical models of AGB stars with
  dust-driven winds}}.
\bibinfo{journal}{{\em \aap}} \bibinfo{volume}{669}, \bibinfo{eid}{A155}.
  \bibinfo{doi}{\doi{10.1051/0004-6361/202244992}}.
\eprint{2301.11836}.

\bibtype{Article}%
\bibitem[{Garc{\'\i}a P{\'e}rez} et al.(2016)]{2016AJ....151..144G}
\bibinfo{author}{{Garc{\'\i}a P{\'e}rez} AE}, \bibinfo{author}{{Allende Prieto}
  C}, \bibinfo{author}{{Holtzman} JA}, \bibinfo{author}{{Shetrone} M},
  \bibinfo{author}{{M{\'e}sz{\'a}ros} S}, \bibinfo{author}{{Bizyaev} D},
  \bibinfo{author}{{Carrera} R}, \bibinfo{author}{{Cunha} K},
  \bibinfo{author}{{Garc{\'\i}a-Hern{\'a}ndez} DA}, \bibinfo{author}{{Johnson}
  JA}, \bibinfo{author}{{Majewski} SR}, \bibinfo{author}{{Nidever} DL},
  \bibinfo{author}{{Schiavon} RP}, \bibinfo{author}{{Shane} N},
  \bibinfo{author}{{Smith} VV}, \bibinfo{author}{{Sobeck} J},
  \bibinfo{author}{{Troup} N}, \bibinfo{author}{{Zamora} O},
  \bibinfo{author}{{Weinberg} DH}, \bibinfo{author}{{Bovy} J},
  \bibinfo{author}{{Eisenstein} DJ}, \bibinfo{author}{{Feuillet} D},
  \bibinfo{author}{{Frinchaboy} PM}, \bibinfo{author}{{Hayden} MR},
  \bibinfo{author}{{Hearty} FR}, \bibinfo{author}{{Nguyen} DC},
  \bibinfo{author}{{O'Connell} RW}, \bibinfo{author}{{Pinsonneault} MH},
  \bibinfo{author}{{Wilson} JC} and  \bibinfo{author}{{Zasowski} G}
  (\bibinfo{year}{2016}), \bibinfo{month}{Jun.}
\bibinfo{title}{{ASPCAP: The APOGEE Stellar Parameter and Chemical Abundances
  Pipeline}}.
\bibinfo{journal}{{\em \aj}} \bibinfo{volume}{151} (\bibinfo{number}{6}),
  \bibinfo{eid}{144}. \bibinfo{doi}{\doi{10.3847/0004-6256/151/6/144}}.
\eprint{1510.07635}.

\bibtype{Book}%
\bibitem[{Gray}(2021)]{Gray2021}
\bibinfo{author}{{Gray} DF} (\bibinfo{year}{2021}).
\bibinfo{title}{{The Observation and Analysis of Stellar Photospheres}}.

\bibtype{Article}%
\bibitem[{Grunhut} et al.(2017)]{Grunhut2017}
\bibinfo{author}{{Grunhut} JH}, \bibinfo{author}{{Wade} GA},
  \bibinfo{author}{{Neiner} C}, \bibinfo{author}{{Oksala} ME},
  \bibinfo{author}{{Petit} V}, \bibinfo{author}{{Alecian} E},
  \bibinfo{author}{{Bohlender} DA}, \bibinfo{author}{{Bouret} JC},
  \bibinfo{author}{{Henrichs} HF}, \bibinfo{author}{{Hussain} GAJ},
  \bibinfo{author}{{Kochukhov} O} and  \bibinfo{author}{{MiMeS Collaboration}}
  (\bibinfo{year}{2017}), \bibinfo{month}{Feb.}
\bibinfo{title}{{The MiMeS survey of Magnetism in Massive Stars: magnetic
  analysis of the O-type stars}}.
\bibinfo{journal}{{\em \mnras}} \bibinfo{volume}{465} (\bibinfo{number}{2}):
  \bibinfo{pages}{2432--2470}. \bibinfo{doi}{\doi{10.1093/mnras/stw2743}}.
\eprint{1610.07895}.

\bibtype{Article}%
\bibitem[{Hawcroft} et al.(2021)]{Hawcroft2021}
\bibinfo{author}{{Hawcroft} C}, \bibinfo{author}{{Sana} H},
  \bibinfo{author}{{Mahy} L}, \bibinfo{author}{{Sundqvist} JO},
  \bibinfo{author}{{Abdul-Masih} M}, \bibinfo{author}{{Bouret} JC},
  \bibinfo{author}{{Brands} SA}, \bibinfo{author}{{de Koter} A},
  \bibinfo{author}{{Driessen} FA} and  \bibinfo{author}{{Puls} J}
  (\bibinfo{year}{2021}), \bibinfo{month}{Nov.}
\bibinfo{title}{{Empirical mass-loss rates and clumping properties of Galactic
  early-type O supergiants}}.
\bibinfo{journal}{{\em \aap}} \bibinfo{volume}{655}, \bibinfo{eid}{A67}.
  \bibinfo{doi}{\doi{10.1051/0004-6361/202140603}}.
\eprint{2108.08340}.

\bibtype{Article}%
\bibitem[{Herrero} et al.(1992)]{Herrero1992}
\bibinfo{author}{{Herrero} A}, \bibinfo{author}{{Kudritzki} RP},
  \bibinfo{author}{{Vilchez} JM}, \bibinfo{author}{{Kunze} D},
  \bibinfo{author}{{Butler} K} and  \bibinfo{author}{{Haser} S}
  (\bibinfo{year}{1992}), \bibinfo{month}{Jul.}
\bibinfo{title}{{Intrinsic parameters of galactic luminous OB stars.}}
\bibinfo{journal}{{\em \aap}} \bibinfo{volume}{261}: \bibinfo{pages}{209--234}.

\bibtype{Article}%
\bibitem[{Hillier} and {Dessart}(2012)]{Hillier2012}
\bibinfo{author}{{Hillier} DJ} and  \bibinfo{author}{{Dessart} L}
  (\bibinfo{year}{2012}), \bibinfo{month}{Jul.}
\bibinfo{title}{{Time-dependent radiative transfer calculations for
  supernovae}}.
\bibinfo{journal}{{\em \mnras}} \bibinfo{volume}{424} (\bibinfo{number}{1}):
  \bibinfo{pages}{252--271}.
  \bibinfo{doi}{\doi{10.1111/j.1365-2966.2012.21192.x}}.
\eprint{1204.0527}.

\bibtype{Article}%
\bibitem[{Holgado} et al.(2018)]{Holgado2018}
\bibinfo{author}{{Holgado} G}, \bibinfo{author}{{Sim{\'o}n-D{\'\i}az} S},
  \bibinfo{author}{{Barb{\'a}} RH}, \bibinfo{author}{{Puls} J},
  \bibinfo{author}{{Herrero} A}, \bibinfo{author}{{Castro} N},
  \bibinfo{author}{{Garcia} M}, \bibinfo{author}{{Ma{\'\i}z Apell{\'a}niz} J},
  \bibinfo{author}{{Negueruela} I} and
  \bibinfo{author}{{Sab{\'\i}n-Sanjuli{\'a}n} C} (\bibinfo{year}{2018}),
  \bibinfo{month}{Jun.}
\bibinfo{title}{{The IACOB project. V. Spectroscopic parameters of the O-type
  stars in the modern grid of standards for spectral classification}}.
\bibinfo{journal}{{\em \aap}} \bibinfo{volume}{613}, \bibinfo{eid}{A65}.
  \bibinfo{doi}{\doi{10.1051/0004-6361/201731543}}.
\eprint{1711.10043}.

\bibtype{Article}%
\bibitem[{Holweger} and {Mueller}(1974)]{1974SoPh...39...19H}
\bibinfo{author}{{Holweger} H} and  \bibinfo{author}{{Mueller} EA}
  (\bibinfo{year}{1974}), \bibinfo{month}{Nov.}
\bibinfo{title}{{The Photospheric Barium Spectrum: Solar Abundance and
  Collision Broadening of Ba II Lines by Hydrogen}}.
\bibinfo{journal}{{\em \solphys}} \bibinfo{volume}{39} (\bibinfo{number}{1}):
  \bibinfo{pages}{19--30}. \bibinfo{doi}{\doi{10.1007/BF00154968}}.

\bibtype{Book}%
\bibitem[{Hubeny} and {Mihalas}(2015)]{Hubeny2015}
\bibinfo{author}{{Hubeny} I} and  \bibinfo{author}{{Mihalas} D}
  (\bibinfo{year}{2015}).
\bibinfo{title}{{Theory of Stellar Atmospheres. An Introduction to
  Astrophysical Non-equilibrium Quantitative Spectroscopic Analysis}}.

\bibtype{Article}%
\bibitem[{Hummer}(1963)]{Hummer1963}
\bibinfo{author}{{Hummer} DG} (\bibinfo{year}{1963}), \bibinfo{month}{Jan.}
\bibinfo{title}{{The ionization structure of planetary nebulae, II. Collisional
  cooling of pure hydrogen nebulae}}.
\bibinfo{journal}{{\em \mnras}} \bibinfo{volume}{125}: \bibinfo{pages}{461}.
  \bibinfo{doi}{\doi{10.1093/mnras/125.5.461}}.

\bibtype{Article}%
\bibitem[{Hummer} and {Rybicki}(1971)]{HR1971}
\bibinfo{author}{{Hummer} DG} and  \bibinfo{author}{{Rybicki} GB}
  (\bibinfo{year}{1971}), \bibinfo{month}{Jan.}
\bibinfo{title}{{Radiative transfer in spherically symmetric systems. The
  conservative grey case}}.
\bibinfo{journal}{{\em \mnras}} \bibinfo{volume}{152}: \bibinfo{pages}{1}.
  \bibinfo{doi}{\doi{10.1093/mnras/152.1.1}}.

\bibtype{Article}%
\bibitem[{Kee} et al.(2021)]{Kee2021}
\bibinfo{author}{{Kee} ND}, \bibinfo{author}{{Sundqvist} JO},
  \bibinfo{author}{{Decin} L}, \bibinfo{author}{{de Koter} A} and
  \bibinfo{author}{{Sana} H} (\bibinfo{year}{2021}), \bibinfo{month}{Feb.}
\bibinfo{title}{{Analytic, dust-independent mass-loss rates for red supergiant
  winds initiated by turbulent pressure}}.
\bibinfo{journal}{{\em \aap}} \bibinfo{volume}{646}, \bibinfo{eid}{A180}.
  \bibinfo{doi}{\doi{10.1051/0004-6361/202039224}}.
\eprint{2101.03070}.

\bibtype{Inproceedings}%
\bibitem[{Koesterke}(2009)]{Koesterke2009}
\bibinfo{author}{{Koesterke} L} (\bibinfo{year}{2009}), \bibinfo{month}{Sep.},
  \bibinfo{title}{{Quantitative Spectroscopy in 3D}}, \bibinfo{editor}{{Hubeny}
  I}, \bibinfo{editor}{{Stone} JM}, \bibinfo{editor}{{MacGregor} K} and
  \bibinfo{editor}{{Werner} K}, (Eds.), \bibinfo{booktitle}{Recent Directions
  in Astrophysical Quantitative Spectroscopy and Radiation Hydrodynamics},
  \bibinfo{series}{American Institute of Physics Conference Series},
  \bibinfo{volume}{1171}, \bibinfo{publisher}{AIP},  \bibinfo{pages}{73--84}.

\bibtype{Article}%
\bibitem[{Kub{\'a}t} et al.(1999)]{Kubat1999}
\bibinfo{author}{{Kub{\'a}t} J}, \bibinfo{author}{{Puls} J} and
  \bibinfo{author}{{Pauldrach} AWA} (\bibinfo{year}{1999}),
  \bibinfo{month}{Jan.}
\bibinfo{title}{{Thermal balance of electrons in calculations of model stellar
  atmospheres}}.
\bibinfo{journal}{{\em \aap}} \bibinfo{volume}{341}: \bibinfo{pages}{587--594}.

\bibtype{Article}%
\bibitem[{Lucy}(1971)]{Lucy1971}
\bibinfo{author}{{Lucy} LB} (\bibinfo{year}{1971}), \bibinfo{month}{Jan.}
\bibinfo{title}{{The Formation of Resonance Lines in Extended and Expanding
  Atmospheres}}.
\bibinfo{journal}{{\em \apj}} \bibinfo{volume}{163}: \bibinfo{pages}{95}.
  \bibinfo{doi}{\doi{10.1086/150748}}.

\bibtype{Article}%
\bibitem[{Lucy} and {Solomon}(1970)]{Lucy1970}
\bibinfo{author}{{Lucy} LB} and  \bibinfo{author}{{Solomon} PM}
  (\bibinfo{year}{1970}), \bibinfo{month}{Mar.}
\bibinfo{title}{{Mass Loss by Hot Stars}}.
\bibinfo{journal}{{\em \apj}} \bibinfo{volume}{159}: \bibinfo{pages}{879}.
  \bibinfo{doi}{\doi{10.1086/150365}}.

\bibtype{Article}%
\bibitem[{Ludwig} et al.(2009)]{2009MmSAI..80..711L}
\bibinfo{author}{{Ludwig} HG}, \bibinfo{author}{{Caffau} E},
  \bibinfo{author}{{Steffen} M}, \bibinfo{author}{{Freytag} B},
  \bibinfo{author}{{Bonifacio} P} and  \bibinfo{author}{{Ku{\v{c}}inskas} A}
  (\bibinfo{year}{2009}), \bibinfo{month}{Jan.}
\bibinfo{title}{{The CIFIST 3D model atmosphere grid.}}
\bibinfo{journal}{{\em \memsai}} \bibinfo{volume}{80}: \bibinfo{pages}{711}.
  \bibinfo{doi}{\doi{10.48550/arXiv.0908.4496}}.
\eprint{0908.4496}.

\bibtype{Article}%
\bibitem[{Maeder}(1999)]{Maeder1999}
\bibinfo{author}{{Maeder} A} (\bibinfo{year}{1999}), \bibinfo{month}{Jul.}
\bibinfo{title}{{Stellar evolution with rotation IV: von Zeipel's theorem and
  anisotropic losses of mass and angular momentum}}.
\bibinfo{journal}{{\em \aap}} \bibinfo{volume}{347}: \bibinfo{pages}{185--193}.

\bibtype{Article}%
\bibitem[{Maeder} and {Meynet}(2000)]{MM2000}
\bibinfo{author}{{Maeder} A} and  \bibinfo{author}{{Meynet} G}
  (\bibinfo{year}{2000}), \bibinfo{month}{Sep.}
\bibinfo{title}{{Stellar evolution with rotation. VI. The Eddington and Omega
  -limits, the rotational mass loss for OB and LBV stars}}.
\bibinfo{journal}{{\em \aap}} \bibinfo{volume}{361}: \bibinfo{pages}{159--166}.
  \bibinfo{doi}{\doi{10.48550/arXiv.astro-ph/0006405}}.
\eprint{astro-ph/0006405}.

\bibtype{Article}%
\bibitem[{Magic} et al.(2013)]{2013A&A...557A..26M}
\bibinfo{author}{{Magic} Z}, \bibinfo{author}{{Collet} R},
  \bibinfo{author}{{Asplund} M}, \bibinfo{author}{{Trampedach} R},
  \bibinfo{author}{{Hayek} W}, \bibinfo{author}{{Chiavassa} A},
  \bibinfo{author}{{Stein} RF} and  \bibinfo{author}{{Nordlund} {\r{A}}}
  (\bibinfo{year}{2013}), \bibinfo{month}{Sep.}
\bibinfo{title}{{The Stagger-grid: A grid of 3D stellar atmosphere models. I.
  Methods and general properties}}.
\bibinfo{journal}{{\em \aap}} \bibinfo{volume}{557}, \bibinfo{eid}{A26}.
  \bibinfo{doi}{\doi{10.1051/0004-6361/201321274}}.
\eprint{1302.2621}.

\bibtype{Article}%
\bibitem[{M{\'e}sz{\'a}ros} et al.(2012)]{2012AJ....144..120M}
\bibinfo{author}{{M{\'e}sz{\'a}ros} S}, \bibinfo{author}{{Allende Prieto} C},
  \bibinfo{author}{{Edvardsson} B}, \bibinfo{author}{{Castelli} F},
  \bibinfo{author}{{Garc{\'\i}a P{\'e}rez} AE}, \bibinfo{author}{{Gustafsson}
  B}, \bibinfo{author}{{Majewski} SR}, \bibinfo{author}{{Plez} B},
  \bibinfo{author}{{Schiavon} R}, \bibinfo{author}{{Shetrone} M} and
  \bibinfo{author}{{de Vicente} A} (\bibinfo{year}{2012}),
  \bibinfo{month}{Oct.}
\bibinfo{title}{{New ATLAS9 and MARCS Model Atmosphere Grids for the Apache
  Point Observatory Galactic Evolution Experiment (APOGEE)}}.
\bibinfo{journal}{{\em \aj}} \bibinfo{volume}{144} (\bibinfo{number}{4}),
  \bibinfo{eid}{120}. \bibinfo{doi}{\doi{10.1088/0004-6256/144/4/120}}.
\eprint{1208.1916}.

\bibtype{Article}%
\bibitem[{Michaud}(1970)]{Michaud1970}
\bibinfo{author}{{Michaud} G} (\bibinfo{year}{1970}), \bibinfo{month}{May}.
\bibinfo{title}{{Diffusion Processes in Peculiar a Stars}}.
\bibinfo{journal}{{\em \apj}} \bibinfo{volume}{160}: \bibinfo{pages}{641}.
  \bibinfo{doi}{\doi{10.1086/150459}}.

\bibtype{Article}%
\bibitem[{Michaud} et al.(2011)]{Michaud2011}
\bibinfo{author}{{Michaud} G}, \bibinfo{author}{{Richer} J} and
  \bibinfo{author}{{Richard} O} (\bibinfo{year}{2011}), \bibinfo{month}{May}.
\bibinfo{title}{{Horizontal branch evolution, metallicity, and sdB stars}}.
\bibinfo{journal}{{\em \aap}} \bibinfo{volume}{529}, \bibinfo{eid}{A60}.
  \bibinfo{doi}{\doi{10.1051/0004-6361/201015997}}.
\eprint{1102.1969}.

\bibtype{Book}%
\bibitem[{Mihalas}(1978)]{Mihalas1978}
\bibinfo{author}{{Mihalas} D} (\bibinfo{year}{1978}).
\bibinfo{title}{{Stellar atmospheres}}.

\bibtype{Article}%
\bibitem[{Moe} and {Di Stefano}(2017)]{Moe2017}
\bibinfo{author}{{Moe} M} and  \bibinfo{author}{{Di Stefano} R}
  (\bibinfo{year}{2017}), \bibinfo{month}{Jun.}
\bibinfo{title}{{Mind Your Ps and Qs: The Interrelation between Period (P) and
  Mass-ratio (Q) Distributions of Binary Stars}}.
\bibinfo{journal}{{\em \apjs}} \bibinfo{volume}{230} (\bibinfo{number}{2}),
  \bibinfo{eid}{15}. \bibinfo{doi}{\doi{10.3847/1538-4365/aa6fb6}}.
\eprint{1606.05347}.

\bibtype{Article}%
\bibitem[{Mokiem} et al.(2005)]{Mokiem2005}
\bibinfo{author}{{Mokiem} MR}, \bibinfo{author}{{de Koter} A},
  \bibinfo{author}{{Puls} J}, \bibinfo{author}{{Herrero} A},
  \bibinfo{author}{{Najarro} F} and  \bibinfo{author}{{Villamariz} MR}
  (\bibinfo{year}{2005}), \bibinfo{month}{Oct.}
\bibinfo{title}{{Spectral analysis of early-type stars using a genetic
  algorithm based fitting method}}.
\bibinfo{journal}{{\em \aap}} \bibinfo{volume}{441} (\bibinfo{number}{2}):
  \bibinfo{pages}{711--733}. \bibinfo{doi}{\doi{10.1051/0004-6361:20053522}}.
\eprint{astro-ph/0506751}.

\bibtype{Article}%
\bibitem[{Nahar}(2016)]{2016NewA...46....1N}
\bibinfo{author}{{Nahar} SN} (\bibinfo{year}{2016}), \bibinfo{month}{Jul.}
\bibinfo{title}{{Photoionization and electron-ion recombination of Ti I}}.
\bibinfo{journal}{{\em \na}} \bibinfo{volume}{46}: \bibinfo{pages}{1--8}.
  \bibinfo{doi}{\doi{10.1016/j.newast.2015.11.003}}.

\bibtype{Article}%
\bibitem[{Ness} et al.(2015)]{2015ApJ...808...16N}
\bibinfo{author}{{Ness} M}, \bibinfo{author}{{Hogg} DW}, \bibinfo{author}{{Rix}
  HW}, \bibinfo{author}{{Ho} AYQ} and  \bibinfo{author}{{Zasowski} G}
  (\bibinfo{year}{2015}), \bibinfo{month}{Jul.}
\bibinfo{title}{{The Cannon: A data-driven approach to Stellar Label
  Determination}}.
\bibinfo{journal}{{\em \apj}} \bibinfo{volume}{808} (\bibinfo{number}{1}),
  \bibinfo{eid}{16}. \bibinfo{doi}{\doi{10.1088/0004-637X/808/1/16}}.
\eprint{1501.07604}.

\bibtype{Article}%
\bibitem[{Ogibalov} and {Shved}(2016)]{2016SoSyR..50..316O}
\bibinfo{author}{{Ogibalov} VP} and  \bibinfo{author}{{Shved} GM}
  (\bibinfo{year}{2016}), \bibinfo{month}{Sep.}
\bibinfo{title}{{An improved model of radiative transfer for the NLTE problem
  in the NIR bands of CO$_{2}$ and CO molecules in the daytime atmosphere of
  Mars. 1. Input data and calculation method}}.
\bibinfo{journal}{{\em Solar System Research}} \bibinfo{volume}{50}
  (\bibinfo{number}{5}): \bibinfo{pages}{316--328}.
  \bibinfo{doi}{\doi{10.1134/S003809461605004X}}.

\bibtype{Article}%
\bibitem[{Oskinova} et al.(2007)]{Oskinova2007}
\bibinfo{author}{{Oskinova} LM}, \bibinfo{author}{{Hamann} WR} and
  \bibinfo{author}{{Feldmeier} A} (\bibinfo{year}{2007}), \bibinfo{month}{Dec.}
\bibinfo{title}{{Neglecting the porosity of hot-star winds can lead to
  underestimating mass-loss rates}}.
\bibinfo{journal}{{\em \aap}} \bibinfo{volume}{476} (\bibinfo{number}{3}):
  \bibinfo{pages}{1331--1340}. \bibinfo{doi}{\doi{10.1051/0004-6361:20066377}}.
\eprint{0704.2390}.

\bibtype{Article}%
\bibitem[{Owocki} and {Rybicki}(1984)]{ORI1984}
\bibinfo{author}{{Owocki} SP} and  \bibinfo{author}{{Rybicki} GB}
  (\bibinfo{year}{1984}), \bibinfo{month}{Sep.}
\bibinfo{title}{{Instabilities in line-driven stellar winds. I. Dependence on
  perturbation wavelength.}}
\bibinfo{journal}{{\em \apj}} \bibinfo{volume}{284}: \bibinfo{pages}{337--350}.
  \bibinfo{doi}{\doi{10.1086/162412}}.

\bibtype{Article}%
\bibitem[{Owocki} et al.(1988)]{OCR1988}
\bibinfo{author}{{Owocki} SP}, \bibinfo{author}{{Castor} JI} and
  \bibinfo{author}{{Rybicki} GB} (\bibinfo{year}{1988}), \bibinfo{month}{Dec.}
\bibinfo{title}{{Time-dependent Models of Radiatively Driven Stellar Winds. I.
  Nonlinear Evolution of Instabilities for a Pure Absorption Model}}.
\bibinfo{journal}{{\em \apj}} \bibinfo{volume}{335}: \bibinfo{pages}{914}.
  \bibinfo{doi}{\doi{10.1086/166977}}.

\bibtype{Article}%
\bibitem[{Owocki} et al.(2004)]{Owocki2004}
\bibinfo{author}{{Owocki} SP}, \bibinfo{author}{{Gayley} KG} and
  \bibinfo{author}{{Shaviv} NJ} (\bibinfo{year}{2004}), \bibinfo{month}{Nov.}
\bibinfo{title}{{A Porosity-Length Formalism for Photon-Tiring-limited Mass
  Loss from Stars above the Eddington Limit}}.
\bibinfo{journal}{{\em \apj}} \bibinfo{volume}{616} (\bibinfo{number}{1}):
  \bibinfo{pages}{525--541}. \bibinfo{doi}{\doi{10.1086/424910}}.
\eprint{astro-ph/0409573}.

\bibtype{Article}%
\bibitem[{Parker}(1960)]{Parker1960}
\bibinfo{author}{{Parker} EN} (\bibinfo{year}{1960}), \bibinfo{month}{Nov.}
\bibinfo{title}{{The Hydrodynamic Theory of Solar Corpuscular Radiation and
  Stellar Winds.}}
\bibinfo{journal}{{\em \apj}} \bibinfo{volume}{132}: \bibinfo{pages}{821}.
  \bibinfo{doi}{\doi{10.1086/146985}}.

\bibtype{Phdthesis}%
\bibitem[{Payne}(1925)]{Payne1925}
\bibinfo{author}{{Payne} CH} (\bibinfo{year}{1925}), \bibinfo{month}{Jan.}
\bibinfo{title}{{Stellar Atmospheres; a Contribution to the Observational Study
  of High Temperature in the Reversing Layers of Stars.}}
\bibinfo{comment}{Ph.D. thesis}, \bibinfo{school}{RADCLIFFE COLLEGE.}

\bibtype{Article}%
\bibitem[{Popa} et al.(2023)]{2023A&A...670A..25P}
\bibinfo{author}{{Popa} SA}, \bibinfo{author}{{Hoppe} R},
  \bibinfo{author}{{Bergemann} M}, \bibinfo{author}{{Hansen} CJ},
  \bibinfo{author}{{Plez} B} and  \bibinfo{author}{{Beers} TC}
  (\bibinfo{year}{2023}), \bibinfo{month}{Feb.}
\bibinfo{title}{{Non-local thermodynamic equilibrium analysis of the
  methylidyne radical molecular lines in metal-poor stellar atmospheres}}.
\bibinfo{journal}{{\em \aap}} \bibinfo{volume}{670}, \bibinfo{eid}{A25}.
  \bibinfo{doi}{\doi{10.1051/0004-6361/202245503}}.
\eprint{2212.06517}.

\bibtype{Article}%
\bibitem[Prandtl(1925)]{Prandtl1925}
\bibinfo{author}{Prandtl L} (\bibinfo{year}{1925}).
\bibinfo{title}{{7. Bericht {\"u}ber Untersuchungen zur ausgebildeten
  Turbulenz}}.
\bibinfo{journal}{{\em ZAMM-Journal of Applied Mathematics and
  Mechanics/Zeitschrift f{\"u}r Angewandte Mathematik und Mechanik}}
  \bibinfo{volume}{5} (\bibinfo{number}{2}): \bibinfo{pages}{136--139}.

\bibtype{Article}%
\bibitem[{Pr{\v{s}}a} et al.(2016)]{Prsa2016}
\bibinfo{author}{{Pr{\v{s}}a} A}, \bibinfo{author}{{Conroy} KE},
  \bibinfo{author}{{Horvat} M}, \bibinfo{author}{{Pablo} H},
  \bibinfo{author}{{Kochoska} A}, \bibinfo{author}{{Bloemen} S},
  \bibinfo{author}{{Giammarco} J}, \bibinfo{author}{{Hambleton} KM} and
  \bibinfo{author}{{Degroote} P} (\bibinfo{year}{2016}), \bibinfo{month}{Dec.}
\bibinfo{title}{{Physics Of Eclipsing Binaries. II. Toward the Increased Model
  Fidelity}}.
\bibinfo{journal}{{\em \apjs}} \bibinfo{volume}{227} (\bibinfo{number}{2}),
  \bibinfo{eid}{29}. \bibinfo{doi}{\doi{10.3847/1538-4365/227/2/29}}.
\eprint{1609.08135}.

\bibtype{Article}%
\bibitem[{Puls}(2009)]{Puls2009}
\bibinfo{author}{{Puls} J} (\bibinfo{year}{2009}), \bibinfo{month}{Jul.}
\bibinfo{title}{{Modeling the atmospheres of massive stars}}.
\bibinfo{journal}{{\em Communications in Asteroseismology}}
  \bibinfo{volume}{158}: \bibinfo{pages}{113}.

\bibtype{Article}%
\bibitem[{Puls} et al.(2008)]{Puls2008}
\bibinfo{author}{{Puls} J}, \bibinfo{author}{{Vink} JS} and
  \bibinfo{author}{{Najarro} F} (\bibinfo{year}{2008}), \bibinfo{month}{Dec.}
\bibinfo{title}{{Mass loss from hot massive stars}}.
\bibinfo{journal}{{\em \aapr}} \bibinfo{volume}{16} (\bibinfo{number}{3-4}):
  \bibinfo{pages}{209--325}. \bibinfo{doi}{\doi{10.1007/s00159-008-0015-8}}.
\eprint{0811.0487}.

\bibtype{Article}%
\bibitem[{Puls} et al.(2020)]{Puls2020}
\bibinfo{author}{{Puls} J}, \bibinfo{author}{{Najarro} F},
  \bibinfo{author}{{Sundqvist} JO} and  \bibinfo{author}{{Sen} K}
  (\bibinfo{year}{2020}), \bibinfo{month}{Oct.}
\bibinfo{title}{{Atmospheric NLTE models for the spectroscopic analysis of blue
  stars with winds. V. Complete comoving frame transfer, and updated modeling
  of X-ray emission}}.
\bibinfo{journal}{{\em \aap}} \bibinfo{volume}{642}, \bibinfo{eid}{A172}.
  \bibinfo{doi}{\doi{10.1051/0004-6361/202038464}}.
\eprint{2011.02310}.

\bibtype{Article}%
\bibitem[{Repolust} et al.(2004)]{Repolust2004}
\bibinfo{author}{{Repolust} T}, \bibinfo{author}{{Puls} J} and
  \bibinfo{author}{{Herrero} A} (\bibinfo{year}{2004}), \bibinfo{month}{Feb.}
\bibinfo{title}{{Stellar and wind parameters of Galactic O-stars. The influence
  of line-blocking/blanketing}}.
\bibinfo{journal}{{\em \aap}} \bibinfo{volume}{415}: \bibinfo{pages}{349--376}.
  \bibinfo{doi}{\doi{10.1051/0004-6361:20034594}}.

\bibtype{Article}%
\bibitem[{Rivero Gonz{\'a}lez} et al.(2012)]{Rivero2012}
\bibinfo{author}{{Rivero Gonz{\'a}lez} JG}, \bibinfo{author}{{Puls} J},
  \bibinfo{author}{{Najarro} F} and  \bibinfo{author}{{Brott} I}
  (\bibinfo{year}{2012}), \bibinfo{month}{Jan.}
\bibinfo{title}{{Nitrogen line spectroscopy of O-stars. II. Surface nitrogen
  abundances for O-stars in the Large Magellanic Cloud}}.
\bibinfo{journal}{{\em \aap}} \bibinfo{volume}{537}, \bibinfo{eid}{A79}.
  \bibinfo{doi}{\doi{10.1051/0004-6361/201117790}}.
\eprint{1110.5148}.

\bibtype{Article}%
\bibitem[{Rodr{\'\i}guez D{\'\i}az} et al.(2024)]{2024arXiv240507872R}
\bibinfo{author}{{Rodr{\'\i}guez D{\'\i}az} LF}, \bibinfo{author}{{Lagae} C},
  \bibinfo{author}{{Amarsi} AM}, \bibinfo{author}{{Bigot} L},
  \bibinfo{author}{{Zhou} Y}, \bibinfo{author}{{Aguirre B{\o}rsen-Koch} V},
  \bibinfo{author}{{Lind} K}, \bibinfo{author}{{Trampedach} R} and
  \bibinfo{author}{{Collet} R} (\bibinfo{year}{2024}), \bibinfo{month}{May}.
\bibinfo{title}{{An extended and refined grid of 3D STAGGER model atmospheres.
  Processed snapshots for stellar spectroscopy}}.
\bibinfo{journal}{{\em arXiv e-prints}} ,
  \bibinfo{eid}{arXiv:2405.07872}\bibinfo{doi}{\doi{10.48550/arXiv.2405.07872}}.
\eprint{2405.07872}.

\bibtype{Article}%
\bibitem[{Rybicki} and {Hummer}(1978)]{RybickiHummer1978}
\bibinfo{author}{{Rybicki} GB} and  \bibinfo{author}{{Hummer} DG}
  (\bibinfo{year}{1978}), \bibinfo{month}{Jan.}
\bibinfo{title}{{A generalization of the Sobolev method for flows with nonlocal
  radiative coupling.}}
\bibinfo{journal}{{\em \apj}} \bibinfo{volume}{219}: \bibinfo{pages}{654--675}.
  \bibinfo{doi}{\doi{10.1086/155826}}.

\bibtype{Article}%
\bibitem[{Sana} et al.(2012)]{Sana2012}
\bibinfo{author}{{Sana} H}, \bibinfo{author}{{de Mink} SE},
  \bibinfo{author}{{de Koter} A}, \bibinfo{author}{{Langer} N},
  \bibinfo{author}{{Evans} CJ}, \bibinfo{author}{{Gieles} M},
  \bibinfo{author}{{Gosset} E}, \bibinfo{author}{{Izzard} RG},
  \bibinfo{author}{{Le Bouquin} JB} and  \bibinfo{author}{{Schneider} FRN}
  (\bibinfo{year}{2012}), \bibinfo{month}{Jul.}
\bibinfo{title}{{Binary Interaction Dominates the Evolution of Massive Stars}}.
\bibinfo{journal}{{\em Science}} \bibinfo{volume}{337}
  (\bibinfo{number}{6093}): \bibinfo{pages}{444}.
  \bibinfo{doi}{\doi{10.1126/science.1223344}}.
\eprint{1207.6397}.

\bibtype{Article}%
\bibitem[{Schneider} et al.(2016)]{Schneider2016}
\bibinfo{author}{{Schneider} FRN}, \bibinfo{author}{{Podsiadlowski} P},
  \bibinfo{author}{{Langer} N}, \bibinfo{author}{{Castro} N} and
  \bibinfo{author}{{Fossati} L} (\bibinfo{year}{2016}), \bibinfo{month}{Apr.}
\bibinfo{title}{{Rejuvenation of stellar mergers and the origin of magnetic
  fields in massive stars}}.
\bibinfo{journal}{{\em \mnras}} \bibinfo{volume}{457} (\bibinfo{number}{3}):
  \bibinfo{pages}{2355--2365}. \bibinfo{doi}{\doi{10.1093/mnras/stw148}}.
\eprint{1601.05084}.

\bibtype{Inproceedings}%
\bibitem[{Schweitzer} et al.(2003)]{2003ASPC..288..339S}
\bibinfo{author}{{Schweitzer} A}, \bibinfo{author}{{Hauschildt} PH},
  \bibinfo{author}{{Baron} E} and  \bibinfo{author}{{Allard} F}
  (\bibinfo{year}{2003}), \bibinfo{month}{Jan.}, \bibinfo{title}{{Using
  Superlevels to Calculate Molecular NLTE Problems}}, \bibinfo{editor}{{Hubeny}
  I}, \bibinfo{editor}{{Mihalas} D} and  \bibinfo{editor}{{Werner} K}, (Eds.),
  \bibinfo{booktitle}{Stellar Atmosphere Modeling},
  \bibinfo{series}{Astronomical Society of the Pacific Conference Series},
  \bibinfo{volume}{288}, pp. \bibinfo{pages}{339}.

\bibtype{Article}%
\bibitem[{Shaviv}(1998)]{Shaviv1998}
\bibinfo{author}{{Shaviv} NJ} (\bibinfo{year}{1998}), \bibinfo{month}{Feb.}
\bibinfo{title}{{The Eddington Luminosity Limit for Multiphased Media}}.
\bibinfo{journal}{{\em \apjl}} \bibinfo{volume}{494} (\bibinfo{number}{2}):
  \bibinfo{pages}{L193--L197}. \bibinfo{doi}{\doi{10.1086/311182}}.

\bibtype{incollection}%
\bibitem[{Sim{\'o}n-D{\'\i}az}(2020)]{SSD2020}
\bibinfo{author}{{Sim{\'o}n-D{\'\i}az} S} (\bibinfo{year}{2020}),
  \bibinfo{title}{{A Modern Guide to Quantitative Spectroscopy of Massive OB
  Stars}}, \bibinfo{booktitle}{Reviews in Frontiers of Modern Astrophysics;
  From Space Debris to Cosmology},  \bibinfo{pages}{155--187}.

\bibtype{Article}%
\bibitem[{Sim{\'o}n-D{\'\i}az} and {Herrero}(2007)]{SSD2007}
\bibinfo{author}{{Sim{\'o}n-D{\'\i}az} S} and  \bibinfo{author}{{Herrero} A}
  (\bibinfo{year}{2007}), \bibinfo{month}{Jun.}
\bibinfo{title}{{Fourier method of determining the rotational velocities in OB
  stars}}.
\bibinfo{journal}{{\em \aap}} \bibinfo{volume}{468} (\bibinfo{number}{3}):
  \bibinfo{pages}{1063--1073}. \bibinfo{doi}{\doi{10.1051/0004-6361:20066060}}.
\eprint{astro-ph/0703216}.

\bibtype{Article}%
\bibitem[{Smith} and {Owocki}(2006)]{Smith2006}
\bibinfo{author}{{Smith} N} and  \bibinfo{author}{{Owocki} SP}
  (\bibinfo{year}{2006}), \bibinfo{month}{Jul.}
\bibinfo{title}{{On the Role of Continuum-driven Eruptions in the Evolution of
  Very Massive Stars and Population III Stars}}.
\bibinfo{journal}{{\em \apjl}} \bibinfo{volume}{645} (\bibinfo{number}{1}):
  \bibinfo{pages}{L45--L48}. \bibinfo{doi}{\doi{10.1086/506523}}.
\eprint{astro-ph/0606174}.

\bibtype{Book}%
\bibitem[{Sobolev}(1960)]{Sobolev1960}
\bibinfo{author}{{Sobolev} VV} (\bibinfo{year}{1960}).
\bibinfo{title}{{Moving Envelopes of Stars}}.
\bibinfo{doi}{\doi{10.4159/harvard.9780674864658}}.

\bibtype{Article}%
\bibitem[{Steffen} and {Freytag}(1991)]{1991RvMA....4...43S}
\bibinfo{author}{{Steffen} M} and  \bibinfo{author}{{Freytag} B}
  (\bibinfo{year}{1991}), \bibinfo{month}{Jan.}
\bibinfo{title}{{Hydrodynamics of the Solar Photosphere: Model Calculations and
  Spectroscopic Observations.}}
\bibinfo{journal}{{\em Reviews in Modern Astronomy}} \bibinfo{volume}{4}:
  \bibinfo{pages}{43--60}. \bibinfo{doi}{\doi{10.1007/978-3-642-76750-0_3}}.

\bibtype{Article}%
\bibitem[{Stein} and {Nordlund}(1989)]{1989ApJ...342L..95S}
\bibinfo{author}{{Stein} RF} and  \bibinfo{author}{{Nordlund} A}
  (\bibinfo{year}{1989}), \bibinfo{month}{Jul.}
\bibinfo{title}{{Topology of Convection beneath the Solar Surface}}.
\bibinfo{journal}{{\em \apjl}} \bibinfo{volume}{342}: \bibinfo{pages}{L95}.
  \bibinfo{doi}{\doi{10.1086/185493}}.

\bibtype{Article}%
\bibitem[{Sundqvist} and {Puls}(2018)]{SP2018}
\bibinfo{author}{{Sundqvist} JO} and  \bibinfo{author}{{Puls} J}
  (\bibinfo{year}{2018}), \bibinfo{month}{Nov.}
\bibinfo{title}{{Atmospheric NLTE models for the spectroscopic analysis of blue
  stars with winds. IV. Porosity in physical and velocity space}}.
\bibinfo{journal}{{\em \aap}} \bibinfo{volume}{619}, \bibinfo{eid}{A59}.
  \bibinfo{doi}{\doi{10.1051/0004-6361/201832993}}.
\eprint{1805.11010}.

\bibtype{Article}%
\bibitem[{Sundqvist} et al.(2018)]{Sundqvist2018}
\bibinfo{author}{{Sundqvist} JO}, \bibinfo{author}{{Owocki} SP} and
  \bibinfo{author}{{Puls} J} (\bibinfo{year}{2018}), \bibinfo{month}{Mar.}
\bibinfo{title}{{2D wind clumping in hot, massive stars from hydrodynamical
  line-driven instability simulations using a pseudo-planar approach}}.
\bibinfo{journal}{{\em \aap}} \bibinfo{volume}{611}, \bibinfo{eid}{A17}.
  \bibinfo{doi}{\doi{10.1051/0004-6361/201731718}}.
\eprint{1710.07780}.

\bibtype{Article}%
\bibitem[{Tennyson} and {Yurchenko}(2018)]{2018Atoms...6...26T}
\bibinfo{author}{{Tennyson} J} and  \bibinfo{author}{{Yurchenko} SN}
  (\bibinfo{year}{2018}), \bibinfo{month}{May}.
\bibinfo{title}{{The ExoMol Atlas of Molecular Opacities}}.
\bibinfo{journal}{{\em Atoms}} \bibinfo{volume}{6} (\bibinfo{number}{2}),
  \bibinfo{eid}{26}. \bibinfo{doi}{\doi{10.3390/atoms6020026}}.
\eprint{1805.03711}.

\bibtype{Article}%
\bibitem[{Tennyson} et al.(2024)]{2024arXiv240606347T}
\bibinfo{author}{{Tennyson} J}, \bibinfo{author}{{Yurchenko} SN},
  \bibinfo{author}{{Zhang} J}, \bibinfo{author}{{Bowesman} CA},
  \bibinfo{author}{{Brady} RP}, \bibinfo{author}{{Buldyreva} J},
  \bibinfo{author}{{Chubb} KL}, \bibinfo{author}{{Gamache} RR},
  \bibinfo{author}{{Gorman} MN}, \bibinfo{author}{{Guest} ER},
  \bibinfo{author}{{Hill} C}, \bibinfo{author}{{Kefala} K},
  \bibinfo{author}{{Lynas-Gray} AE}, \bibinfo{author}{{Mellor} TM},
  \bibinfo{author}{{McKemmish} LK}, \bibinfo{author}{{Mitev} GB},
  \bibinfo{author}{{Mizus} II}, \bibinfo{author}{{Owens} A},
  \bibinfo{author}{{Peng} Z}, \bibinfo{author}{{Perri} AN},
  \bibinfo{author}{{Pezzella} M}, \bibinfo{author}{{Polyansky} OL},
  \bibinfo{author}{{Qu} Q}, \bibinfo{author}{{Semenov} M},
  \bibinfo{author}{{Smola} O}, \bibinfo{author}{{Solokov} A},
  \bibinfo{author}{{Somogyi} W}, \bibinfo{author}{{Upadhyay} A},
  \bibinfo{author}{{Wright} SOM} and  \bibinfo{author}{{Zobov} NF}
  (\bibinfo{year}{2024}), \bibinfo{month}{Jun.}
\bibinfo{title}{{The 2024 release of the ExoMol database: molecular line lists
  for exoplanet and other hot atmospheres}}.
\bibinfo{journal}{{\em arXiv e-prints}} ,
  \bibinfo{eid}{arXiv:2406.06347}\bibinfo{doi}{\doi{10.48550/arXiv.2406.06347}}.
\eprint{2406.06347}.

\bibtype{Article}%
\bibitem[{Ting} et al.(2019)]{2019ApJ...879...69T}
\bibinfo{author}{{Ting} YS}, \bibinfo{author}{{Conroy} C},
  \bibinfo{author}{{Rix} HW} and  \bibinfo{author}{{Cargile} P}
  (\bibinfo{year}{2019}), \bibinfo{month}{Jul.}
\bibinfo{title}{{The Payne: Self-consistent ab initio Fitting of Stellar
  Spectra}}.
\bibinfo{journal}{{\em \apj}} \bibinfo{volume}{879} (\bibinfo{number}{2}),
  \bibinfo{eid}{69}. \bibinfo{doi}{\doi{10.3847/1538-4357/ab2331}}.
\eprint{1804.01530}.

\bibtype{Article}%
\bibitem[{ud-Doula} and {Owocki}(2002)]{udDoula2002}
\bibinfo{author}{{ud-Doula} A} and  \bibinfo{author}{{Owocki} SP}
  (\bibinfo{year}{2002}), \bibinfo{month}{Sep.}
\bibinfo{title}{{Dynamical Simulations of Magnetically Channeled Line-driven
  Stellar Winds. I. Isothermal, Nonrotating, Radially Driven Flow}}.
\bibinfo{journal}{{\em \apj}} \bibinfo{volume}{576} (\bibinfo{number}{1}):
  \bibinfo{pages}{413--428}. \bibinfo{doi}{\doi{10.1086/341543}}.
\eprint{astro-ph/0201195}.

\bibtype{Book}%
\bibitem[{Uns\"old}(1955)]{Unsold1955}
\bibinfo{author}{{Uns\"old} A} (\bibinfo{year}{1955}).
\bibinfo{title}{{Physik der Sternatmosph\"aren, mit besonderer
  Ber\"ucksichtigung der Sonne.}}

\bibtype{Article}%
\bibitem[{Vink}(2022)]{Vink2022}
\bibinfo{author}{{Vink} JS} (\bibinfo{year}{2022}), \bibinfo{month}{Aug.}
\bibinfo{title}{{Theory and Diagnostics of Hot Star Mass Loss}}.
\bibinfo{journal}{{\em \araa}} \bibinfo{volume}{60}: \bibinfo{pages}{203--246}.
  \bibinfo{doi}{\doi{10.1146/annurev-astro-052920-094949}}.
\eprint{2109.08164}.

\bibtype{Article}%
\bibitem[{von Zeipel}(1924)]{vonZeipel1924}
\bibinfo{author}{{von Zeipel} H} (\bibinfo{year}{1924}), \bibinfo{month}{Jun.}
\bibinfo{title}{{The radiative equilibrium of a rotating system of gaseous
  masses}}.
\bibinfo{journal}{{\em \mnras}} \bibinfo{volume}{84}:
  \bibinfo{pages}{665--683}. \bibinfo{doi}{\doi{10.1093/mnras/84.9.665}}.

\bibtype{Article}%
\bibitem[{Wade} et al.(2016)]{Wade2016}
\bibinfo{author}{{Wade} GA}, \bibinfo{author}{{Neiner} C},
  \bibinfo{author}{{Alecian} E}, \bibinfo{author}{{Grunhut} JH},
  \bibinfo{author}{{Petit} V}, \bibinfo{author}{{Batz} Bd},
  \bibinfo{author}{{Bohlender} DA}, \bibinfo{author}{{Cohen} DH},
  \bibinfo{author}{{Henrichs} HF}, \bibinfo{author}{{Kochukhov} O},
  \bibinfo{author}{{Landstreet} JD}, \bibinfo{author}{{Manset} N},
  \bibinfo{author}{{Martins} F}, \bibinfo{author}{{Mathis} S},
  \bibinfo{author}{{Oksala} ME}, \bibinfo{author}{{Owocki} SP},
  \bibinfo{author}{{Rivinius} T}, \bibinfo{author}{{Shultz} ME},
  \bibinfo{author}{{Sundqvist} JO}, \bibinfo{author}{{Townsend} RHD},
  \bibinfo{author}{{ud-Doula} A}, \bibinfo{author}{{Bouret} JC},
  \bibinfo{author}{{Braithwaite} J}, \bibinfo{author}{{Briquet} M},
  \bibinfo{author}{{Carciofi} AC}, \bibinfo{author}{{David-Uraz} A},
  \bibinfo{author}{{Folsom} CP}, \bibinfo{author}{{Fullerton} AW},
  \bibinfo{author}{{Leroy} B}, \bibinfo{author}{{Marcolino} WLF},
  \bibinfo{author}{{Moffat} AFJ}, \bibinfo{author}{{Naz{\'e}} Y},
  \bibinfo{author}{{Louis} NS}, \bibinfo{author}{{Auri{\`e}re} M},
  \bibinfo{author}{{Bagnulo} S}, \bibinfo{author}{{Bailey} JD},
  \bibinfo{author}{{Barb{\'a}} RH}, \bibinfo{author}{{Blaz{\`e}re} A},
  \bibinfo{author}{{B{\"o}hm} T}, \bibinfo{author}{{Catala} C},
  \bibinfo{author}{{Donati} JF}, \bibinfo{author}{{Ferrario} L},
  \bibinfo{author}{{Harrington} D}, \bibinfo{author}{{Howarth} ID},
  \bibinfo{author}{{Ignace} R}, \bibinfo{author}{{Kaper} L},
  \bibinfo{author}{{L{\"u}ftinger} T}, \bibinfo{author}{{Prinja} R},
  \bibinfo{author}{{Vink} JS}, \bibinfo{author}{{Weiss} WW} and
  \bibinfo{author}{{Yakunin} I} (\bibinfo{year}{2016}), \bibinfo{month}{Feb.}
\bibinfo{title}{{The MiMeS survey of magnetism in massive stars: introduction
  and overview}}.
\bibinfo{journal}{{\em \mnras}} \bibinfo{volume}{456} (\bibinfo{number}{1}):
  \bibinfo{pages}{2--22}. \bibinfo{doi}{\doi{10.1093/mnras/stv2568}}.
\eprint{1511.08425}.

\bibtype{Article}%
\bibitem[{Werner} and {Husfeld}(1985)]{Werner1985}
\bibinfo{author}{{Werner} K} and  \bibinfo{author}{{Husfeld} D}
  (\bibinfo{year}{1985}), \bibinfo{month}{Jul.}
\bibinfo{title}{{Multi-level non-LTE line formation calculations using
  approximate Lambda-operators}}.
\bibinfo{journal}{{\em \aap}} \bibinfo{volume}{148} (\bibinfo{number}{2}):
  \bibinfo{pages}{417--422}.

\bibtype{Article}%
\bibitem[{Wilson} and {Devinney}(1971)]{Wilson1971}
\bibinfo{author}{{Wilson} RE} and  \bibinfo{author}{{Devinney} EJ}
  (\bibinfo{year}{1971}), \bibinfo{month}{Jun.}
\bibinfo{title}{{Realization of Accurate Close-Binary Light Curves: Application
  to MR Cygni}}.
\bibinfo{journal}{{\em \apj}} \bibinfo{volume}{166}: \bibinfo{pages}{605}.
  \bibinfo{doi}{\doi{10.1086/150986}}.

\bibtype{Article}%
\bibitem[{Younsi} et al.(2012)]{Younsi2012}
\bibinfo{author}{{Younsi} Z}, \bibinfo{author}{{Wu} K} and
  \bibinfo{author}{{Fuerst} SV} (\bibinfo{year}{2012}), \bibinfo{month}{Sep.}
\bibinfo{title}{{General relativistic radiative transfer: formulation and
  emission from structured tori around black holes}}.
\bibinfo{journal}{{\em \aap}} \bibinfo{volume}{545}, \bibinfo{eid}{A13}.
  \bibinfo{doi}{\doi{10.1051/0004-6361/201219599}}.
\eprint{1207.4234}.

\end{thebibliography*}

\end{document}